\documentclass[aps,prc,reprint,showpacs,superscriptaddress,onecolumn,notitlepage]{revtex4-1}

\usepackage{graphicx,color} 
\usepackage{bm}       
\usepackage{amsmath} 
\usepackage[dvipsnames]{xcolor}
\usepackage{ulem}
\usepackage{siunitx}
\usepackage{multirow}
\usepackage{gensymb}
\usepackage{url}
\begin{document}

\newcommand {\nc} {\newcommand}

\newcommand{\vv}[1]{{$\bf {#1}$}}
\newcommand{\vvm}[1]{{\bf {#1}}}
\def\btau{\mbox{\boldmath$\tau$}}

\nc {\IR} [1]{\textcolor{red}{#1}}
\nc {\IB} [1]{\textcolor{blue}{#1}}
\nc {\IP} [1]{\textcolor{magenta}{#1}}
\nc {\IM} [1]{\textcolor{Bittersweet}{#1}}
\nc {\IE} [1]{\textcolor{Plum}{#1}}

\nc{\ninej}[9]{\left\{\begin{array}{ccc} #1 & #2 & #3 \\ #4 & #5 & #6 \\ #7 & #8 & #9 \\ \end{array}\right\}}
\nc{\sixj}[6]{\left\{\begin{array}{ccc} #1 & #2 & #3 \\ #4 & #5 & #6 \\ \end{array}\right\}}
\nc{\threej}[6]{ \left( \begin{array}{ccc} #1 & #2 & #3 \\ #4 & #5 & #6 \\ \end{array} \right) }
\nc{\half}{\frac{1}{2}}
\nc{\numberthis}{\addtocounter{equation}{1}\tag{\theequation}}
\nc{\lla}{\left\langle}
\nc{\rra}{\right\rangle}
\nc{\lrme}{\left|\left|}
\nc{\rrme}{\right|\right|}

\nc{\weppner}{\IB}

\title{Nuclear structure and elastic scattering observables obtained consistently with different NN
interactions} 
\author{R.~B.~Baker}
\affiliation{Institute of Nuclear and Particle Physics, and Department of Physics and Astronomy, Ohio University, Athens, OH 45701, USA}
\author{M.~Burrows}
\affiliation{Department of Physics and Astronomy, Louisiana State University, Baton Rouge, LA 70803, USA}
\author{Ch.~Elster}
\affiliation{Institute of Nuclear and Particle Physics, and Department of Physics and Astronomy, Ohio University, Athens, OH 45701, USA}
\author{P.~Maris}
\affiliation{Department of Physics and Astronomy, Iowa State University, Ames, IA 50011, USA}
\author{G.~Popa}
\affiliation{Institute of Nuclear and Particle Physics, and Department of Physics and Astronomy, Ohio University, Athens, OH 45701, USA}
\author{S.~P.~Weppner}
\affiliation{Natural Sciences, Eckerd College, St. Petersburg, FL 33711, USA}

\date{\today}

\begin{abstract}
Nucleon-nucleon ($NN$) interactions based on chiral effective theories
are commonly used in ab initio calculations of light nuclei.
Here we present a study based on three different $NN$ interactions (up to
next-to-next-to-leading order) for which structure and elastic proton scattering observables
are consistently calculated for $^4$He, $^{12}$C, and $^{16}$O.
The interactions are compared at the two-body level in terms of Wolfenstein amplitudes,
and their predictions for ground state energies, point-proton radii, and charge form factors,
as well as proton elastic scattering observables in the leading-order spectator expansion
in the energy range between 65 and 160 MeV projectile energy are presented.
To gain further insight into differences visible in elastic scattering observables,
we investigate the behavior of the calculated effective nucleon-nucleus interactions
for the $^{12}$C nucleus based on the different $NN$ interactions.


\end{abstract}


\maketitle

\section{Introduction}
\label{sec:intro}

Over the last decade developments of the nucleon-nucleon ($NN$) and three-nucleon ($3N$) interactions
derived from chiral effective field theory have yielded major
progress~\cite{EntemM03,Epelbaum06,Epelbaum:2008ga,Epelbaum:2014sza,Epelbaum:2014efa,Reinert:2017usi,Machleidt:2011zz,Entem:2017gor}.
They, together with the utilization of massively parallel computing resources (e.g.,
see~\cite{LangrDDLT19,LangrDDT18,SHAO20181,CPE:CPE3129,Jung:2013:EFO}), have placed {\it ab initio}
large-scale simulations at the frontier of nuclear structure and reaction explorations. 
Among other
successful many-body theories, the {\it ab initio} no-core shell model (NCSM) approach (see, e.g.,
\cite{Navratil:2000ww,Roth:2007sv,Barrett:2013nh,Binder:2018pgl}), has over
the last decade taken center stage in the development of microscopic tools for studying the
structure of atomic nuclei up to $A\simeq 16$. 

Following the developments in {\it ab initio} structure theory, rigorous calculations of effective
folding nucleon-nucleus ($NA$) interactions for elastic scattering of protons or neutrons from 
nuclei in the same mass region were developed~\cite{Burrows:2018ggt,Burrows:2020qvu,Baker:2021izp,Baker:2021iqy,Gennari:2017yez,Vorabbi:2021kho,Vorabbi:2020cgf,Arellano:2022tsi}
 based on the leading-order in the spectator expansion of multiple scattering theory~\cite{Siciliano:1977zz,Ernst:1977gb,Tandy:1980zz}.
Here the nuclear one-body densities that are required for the folding with $NN$ scattering amplitudes can
be based on the same chiral $NN$ interaction. This development not only allows one to consider $NA$ effective
interaction derived in leading-order of the spectator expansion as {\it ab initio}, but also allows
an investigation of effects of truncation uncertainties  in
the chiral expansion on $NA$ scattering observables in a similar fashion as already successfully
performed in $NN$ scattering (see e.g.~\cite{Furnstahl:2015rha,Melendez:2017phj,Melendez:2019izc}),
nucleon-deuteron scattering~\cite{Epelbaum:2019zqc}, or structure observables for light
nuclei~\cite{Binder:2018pgl,Maris:2020qne}.

To date there is no unambiguous description of the strong interaction between
nucleons based on effective field theories. Most strikingly, it is not fully understood why some descriptions of
the nuclear interaction based on this framework and calibrated to reproduce similar data in the
few-nucleon sector do better than others in predicting fundamental nuclear quantities such as
 binding energies, radii, decays, or cross sections. Though this work will not be able to answer those
questions,  our aim is to illustrate those differences by looking at structure and elastic
scattering observables simultaneously. 
For our study we choose three different $NN$ interactions based on chiral effective theory at
next-to-next-to-leading order, which are
calibrated to $NN$ data. 

The construction of the effective $NA$ interaction in leading order in the spectator expansion is
briefly reviewed in Section~\ref{sec:formal} to introduce how the $NN$ interaction enters
the elastic scattering calculations. Then the three different interactions based on
chiral effective theory, namely the NNLO$_{\rm{opt}}$~\cite{Ekstrom13} and the
EKM~\cite{Epelbaum:2014sza,Epelbaum:2014efa} chiral interactions as well as  the Daejeon16
potential~\cite{Shirokov:2016ead}, are introduced together with $NN$ amplitudes which are relevant for 
the construction of the effective $NA$ interaction.   Section~\ref{sec:stresults} considers structure
observables for $^4$He, $^{12}$C, and $^{16}$O predicted from those interactions, as well as the charge
form factors, as experimentally extracted from elastic electron-nucleus scattering.
In Section~\ref{sec:scresults} proton-nucleus elastic scattering from those nuclei in the energy range
from 65 to 160~MeV laboratory kinetic energy are presented.  
Effective $NA$ interactions are in general nonlocal as well as energy dependent.
In order to gain insight into their structure and enable a connection to phenomenological local optical
potentials we extract   in Section~\ref{sec:loptpots} the local part and parameterize it with
traditional Woods-Saxon forms in the appendix. We conclude in Section~\ref{sec:conclusions}.


\section{Ab initio effective NA interactions}
\label{sec:formal}

Calculating elastic nucleon-nucleus scattering observables in an {\it ab initio} fashion
requires the interaction between the projectile and the nucleons in the target, as well as the interaction between the nucleons within the target. A multiple scattering expansion can organize the interactions between the projectile and the nucleons in the target according to the number of active nucleons participating in the reaction, as is done 
e.g. in the spectator expansion~\cite{Siciliano:1977zz,PICKLESIMER1983207,Baker:2023uzx}. The structure of
the target nucleus is calculated employing {\it ab initio} many-body methods. For this work we are using the
no-core shell model (NCSM)~\cite{Barrett:2013nh}, which is well-suited for light nuclei; more details about the nuclear structure calculation are given in Section~\ref{sec:stresults} below.

Up to now the leading order in the spectator expansion, which implies two active nucleons, has been evaluated {\it ab initio} in several 
works for elastic scattering of protons (neutrons) from nuclei with a $0^+$ ground state~\cite{Burrows:2020qvu,Gennari:2017yez} as well as
nuclei with $J^\pi\neq 0^+$~\cite{Vorabbi:2021kho}. In this work we concentrate on proton scattering
from light nuclei with  $J^\pi= 0^+$ in leading order in the spectator expansion.
In this case the effective interaction of the projectile proton with a single target nucleon can be written as 
function of the momentum transfer ${\bf q}$ and the average momentum ${\bm{\mathcal{K}}_{NA}}$, where the
subscript $NA$ refers to the nucleon-nucleus ($NA$) frame.
\begin{eqnarray}
\label{eq:1}
\lefteqn{\widehat{U}_{\mathrm{p}}(\bm{q},\bm{\mathcal{K}}_{NA};\epsilon) =} & &  \cr
& & \sum_{\alpha=\mathrm{n,p}} \int d^3{\mathcal{K}} \eta\left( \bm{q}, \bm{\mathcal{K}}, \bm{\mathcal{K}}_{NA} \right) 
A_{\mathrm{p}\alpha}\left( \bm{q}, \frac{1}{2}\left( \frac{A+1}{A}\bm{\mathcal{K}}_{NA} - \bm{\mathcal{K}} \right); \epsilon
\right) \rho_\alpha^{K_s=0} \left(\bm{\mathcal{P}'}, \bm{\mathcal{P}}  \right) \cr  
&+&i (\bm{\sigma^{(0)}}\cdot\hat{\bm{n}}) \sum_{\alpha=\mathrm{n,p}} \int d^3{\mathcal{K}} \eta\left( \bm{q},
\bm{\mathcal{K}}, \bm{\mathcal{K}}_{NA} \right) 
C_{\mathrm{p}\alpha}\left( \bm{q}, \frac{1}{2}\left( \frac{A+1}{A}\bm{\mathcal{K}}_{NA} - \bm{\mathcal{K}} \right);
\epsilon
\right) \rho_\alpha^{K_s=0} \left(\bm{\mathcal{P}'}, \bm{\mathcal{P}}  \right) \cr 
&+&i \sum_{\alpha=\mathrm{n,p}} \int d^3{\mathcal{K}} \eta\left( \bm{q}, \bm{\mathcal{K}}, \bm{\mathcal{K}}_{NA}
\right) C_{\mathrm{p}\alpha} \left( \bm{q}, \frac{1}{2}\left( \frac{A+1}{A}\bm{\mathcal{K}}_{NA} - \bm{\mathcal{K}}
\right); \epsilon \right) S_{n,\alpha} \left(\bm{\mathcal{P}'}, \bm{\mathcal{P}} \right) \cos \beta\cr
&+&i (\bm{\sigma^{(0)}}\cdot\hat{\bm{n}}) \sum_{\alpha=\mathrm{n,p}} \int d^3{\mathcal{K}} \eta\left( \bm{q},
\bm{\mathcal{K}}, \bm{\mathcal{K}}_{NA} \right)  (-i) 
M_{\mathrm{p}\alpha} \left( \bm{q}, \frac{1}{2}\left( \frac{A+1}{A}\bm{\mathcal{K}}_{NA} - \bm{\mathcal{K}}
	\right); \epsilon \right) S_{n,\alpha} \left(\bm{\mathcal{P}'}, \bm{\mathcal{P}}  \right) \cos \beta,
\end{eqnarray}
where the subscript p indicates the projectile being a proton. The energy $\epsilon$ is taken in the impulse
approximation as half of the projectile energy.
The momentum vectors in the problem are given as
\begin{eqnarray}
\label{eq:2}
\bm{q} &=& \bm{p'} - \bm{p} = \bm{k'} - \bm{k}, \cr
\bm{\mathcal{K}} &=& \frac{1}{2} \left(\bm{p'} + \bm{p}\right), \cr
\hat{\bm{n}}&=&\frac{\bm{\mathcal{K}} \times \bm{q}}{\left| \bm{\mathcal{K}} \times
\bm{q}\right|} \cr
\bm{\mathcal{K}_{NA}} &=& \frac{A}{A+1}\left[\left(\bm{k'} + \bm{k}\right) +
      \frac{1}{2} \left(\bm{p'} + \bm{p}\right) \right], \cr
\bm{\mathcal{P}}&=& \bm{\mathcal{K}}+\frac{A-1}{A}\frac{\bm{q}}{2},  \cr
\bm{\mathcal{P'}}&=& \bm{\mathcal{K}}-\frac{A-1}{A}\frac{\bm{q}}{2}  .
\end{eqnarray}
The momentum of the incoming proton is given by $\bm{k}$, its outgoing momentum by $\bm{k'}$, the momentum
transfer by $\bm{q}$, and the average momentum $\bm{\mathcal{K}_{NA}}$. The struck nucleon in the target has an initial momentum $\bm{p}$ and a final momentum $\bm{p'}$.  The two quantities representing the structure of the nucleus are the scalar one-body density
$\rho_\alpha^{K_s=0} \left(\bm{\mathcal{P}'}, \bm{\mathcal{P}}  \right)$ and the
spin-projected momentum distribution $S_{n,\alpha} \left(\bm{\mathcal{P}'}, \bm{\mathcal{P}} \right)$. 
Both distributions are nonlocal and translationally invariant.
Lastly, the term $\cos \beta$ in Eq.~(\ref{eq:1}) comes from projecting $\bm{\hat{n}}$ from the $NN$ frame 
to the $NA$ frame. For further details see Ref.~\cite{Burrows:2020qvu}. The term $\eta\left( \bm{q}, \bm{\mathcal{K}}, \bm{\mathcal{K}_{NA}} \right)$  is the M{\o}ller
factor~\cite{CMoller} describing the transformation from the $NN$ frame to the $NA$ frame.

When calculating $NA$ elastic scattering amplitudes, 
the leading order term of Eq.~(\ref{eq:1}) does not directly enter a Lippmann-Schwinger type integral
equation. To obtain the Watson optical potential $U_p(\bm{q},\bm{\mathcal{K}}_{NA};\epsilon)$, 
an additional integral equation needs to be solved~\cite{Baker:2023uzx,Burrows:2018ggt},
\begin{equation}
\label{eq:watson}
U_{\rm p} = \widehat{U}_{\rm p} - \widehat{U}_{\rm p} G_0 (E) P U_{\rm p},
\end{equation}
where for simplicity the momentum variables are omitted. Here $G_0(E)$ is the free $NA$ propagator and $P$
a projector on the ground state. 

The functions $A_{\mathrm{p}\alpha}$, $C_{\mathrm{p}\alpha}$, and $M_{\mathrm{p}\alpha}$ represent the $NN$ interaction through Wolfenstein
amplitudes (see discussion below). Since the incoming proton can interact
with either a proton or a neutron in the nucleus, the index $\alpha$ indicates the
neutron ($\mathrm{n}$) and proton ($\mathrm{p}$) contributions, which are calculated separately and then summed up.
With respect to the nucleus, the operator $i (\bm{\sigma^{(0)}}\cdot \hat{\bm{n}})$ represents the spin-orbit operator in momentum space of
the projectile. As such, Eq.~(\ref{eq:1}) exhibits the
expected form of an interaction between a spin-$\frac{1}{2}$ projectile and a target nucleus in a $J=0$ 
state \cite{RodbergThaler}.

The effective $NA$ interaction in the leading order spectator expansion is driven by the $NN$ amplitude, which in its
most general form can be parameterized in terms of Wolfenstein
amplitudes~\cite{wolfenstein-ashkin,Fachruddin:2000wv,Golak:2010wz},
\begin{eqnarray}
\label{eq:3}
       \overline{M}(\bm{q},\bm{\mathcal{K}}_{NN},\epsilon)&=&
A(\bm{q},\bm{\mathcal{K}}_{NN},\epsilon)\textbf{1}\otimes\textbf{1} \nonumber \\
       &+& iC(\bm{q},\bm{\mathcal{K}}_{NN},\epsilon)~\left(\bm{\sigma^{(0)}} \cdot \hat{\bm{n}}
\right)\otimes\textbf{1} \nonumber \\
       &+& iC(\bm{q},\bm{\mathcal{K}}_{NN},\epsilon)~\textbf{1}\otimes\left(\bm{\sigma^{(i)}}\cdot
\hat{\bm{n}} \right) \nonumber \\
       &+&
M(\bm{q},\bm{\mathcal{K}}_{NN},\epsilon)(\bm{\sigma^{(0)}}\cdot\hat{\bm{n}})\otimes(\bm{\sigma^{(i)}}\cdot\hat{\bm{n}}) 
\nonumber \\
       &+&
\left[G(\bm{q},\bm{\mathcal{K}}_{NN},\epsilon)-H(\bm{q},\bm{\mathcal{K}}_{NN},\epsilon)\right](\bm{\sigma^{(0)}}\cdot\hat{\bm{q}})\otimes(\bm{\sigma^{(i)}}\cdot\hat{\bm{q}})
\cr
       &+&
\left[G(\bm{q},\bm{\mathcal{K}}_{NN},\epsilon)+H(\bm{q},\bm{\mathcal{K}}_{NN},\epsilon)\right](\bm{\sigma^{(0)}}\cdot\hat{\bm{\mathcal{K}}})\otimes(\bm{\sigma^{(i)}}\cdot\hat{\bm{\mathcal{K}}})
\cr
       &+&
D(\bm{q},\bm{\mathcal{K}}_{NN},\epsilon)\left[(\bm{\sigma^{(0)}}\cdot\hat{\bm{q}})\otimes(\bm{\sigma^{(i)}}\cdot\hat{\bm{\mathcal{K}}})+(\bm{\sigma^{(0)}}\cdot\hat{\bm{\mathcal{K}}})\otimes(\bm{\sigma^{(i)}}\cdot\hat{\bm{q}})\right]~,
\end{eqnarray}
where $\bm{\sigma^{(0)}}$ describes the spin of the projectile, and $\bm{\sigma^{(i)}}$ the spin
of the struck nucleon.
The average momentum in the $NN$ frame is defined as $\bm{\mathcal{K}}_{NN} =
\frac{1}{2} \left(\bm{k'}_{NN} + \bm{k}_{NN}\right)$. The scalar functions $A$, $C$, $M$, $G$, $H$,
and $D$ are referred to as
Wolfenstein amplitudes, and only depend on the scattering momenta and the energy, and are calculated for $np$ and
$pp$ scattering respectively. The amplitude $D(\bm{q},\bm{\mathcal{K}}_{NN},\epsilon)$ vanishes on-shell due to parity invariance. Each term in
Eq.~(\ref{eq:3}) is described by two components, namely a scalar function of two vector momenta and an energy
(for $NN$ scattering this is the c.~m.~energy of the $NN$ system), and the coupling between the operators
of the projectile and the struck nucleon. The Wolfenstein amplitude $A$ sums up all pieces of the central $NN$ force, while $C$ represents all pieces contributing to the spin-orbit force. The amplitudes $M$, $G$, and $H$ sum up tensor force contributions.    

For the struck target nucleon $i$ the expectation values of the of the operators ${\bf 1}$ and the scalar
products of $\bm{\sigma^{(i)}}$ with the linear independent unit vectors $\hat{\bm n}$, $\hat{\bm{q}}$,
and $\bm{\mathcal{K}}_{NN}$ need to be 
calculated for the leading order $NA$ effective interaction of Eq.~(\ref{eq:1}). 
Evaluating the expectation
value of the operator ${\bf 1}$ in the ground state of
the nucleus results in the scalar nonlocal, translationally invariant one-body density
$\rho^{K_s=0}_\alpha$ in Eq.~(\ref{eq:1}). 
This has traditionally and very successfully been used as input
to microscopic or {\it ab initio} calculations of leading order effective
interactions~\cite{Elster:1996xh,Elster:1989en,Burrows:2018ggt,Gennari:2017yez,Arellano:2018jjd}.
The other operators from Eq.~(\ref{eq:2}), namely $(\bm{\sigma^{(i)}} \cdot \hat{\bm n})$,
$(\bm{\sigma^{(i)}} \cdot \hat{\bm q})$, and $(\bm{\sigma^{(i)}} \cdot \hat{\bm {\mathcal{K}}})$
need to also be evaluated for a leading-order {\it ab initio} $NA$ effective interaction, in which the
$NN$ interaction is treated on equal footing in the reaction and structure calculation.
Due to parity invariance arguments only the terms proportional to $(\bm{\sigma^{(i)}} \cdot \hat{\bm n})$ 
do not vanish in the sum over $\alpha$ for $J^\pi= 0^+$ nuclei, leading to spin-projected non-local one-body density $S_{n,\alpha}$ in
Eq.~(\ref{eq:1})~\cite{Burrows:2020qvu,BurrowsM:2020}. Thus the tensor contributions of the $NN$ force
only enter the leading
order effective $NA$ interaction through the Wolfenstein amplitude $M$ as long as elastic scattering is
considered.

For the study of $NA$ observables calculated in leading order of the spectator expansion we choose 
three different chiral $NN$ interactions. One is the optimized chiral $NN$ interaction at the
next-to-next-to-leading order NNLO$_{\rm{opt}}$ from Ref.~\cite{Ekstrom13}. This interaction is
fitted with $\chi^2 \approx 1$ per degree of freedom for laboratory energies up to about 125~MeV. In the
$A$~=~3, 4 nucleon systems the contributions of the 3NFs are smaller than in most other
parameterizations of chiral interactions. As a consequence, nuclear quantities like root-mean-square
radii
and electromagnetic transitions in light and intermediate-mass nuclei can be calculated  reasonably well
without invoking 3NFs \cite{Henderson:2017dqc,LauneyMSSBMDD18}. The second is the chiral $NN$ interaction
from Refs.~\cite{Epelbaum:2014sza,Epelbaum:2014efa} (sometimes referred to as EKM) with a semi-local
cutoff $R=1.0$~fm which we consider up to next-to-next-to-leading order. This interaction has been
employed in Ref.~\cite{Baker:2021iqy} to quantify truncation errors of the chiral EFT in $NA$ observables. 
As third interaction we employ the Daejeon16 potential~\cite{Shirokov:2016ead}, which is based on
the Idaho N3LO chiral interaction~\cite{Entem:2003ft}, which is SRG evolved with a flow parameter
$\lambda$=1.5~fm$^{-1}$ and cast into a harmonic oscillator basis. In addition phase-equivalent transformations are employed in some partial waves to minimize the need for the explicit introduction of three- and higher-body forces~\cite{Shirokov:2016ead}.

All three chiral $NN$ interaction listed above describe the $np$ and $pp$ phase shifts equally well for $NN$
laboratory kinetic energies up about 150 MeV, with differences being in details. For the application in $NA$
scattering the focus may better be directed toward considering the relevant Wolfenstein amplitudes that enter
the $NA$ effective interaction of Eq.~(\ref{eq:1}). Since Wolfenstein amplitudes are complex functions, we
consider in Fig.~\ref{fig1} the squares of the real and imaginary parts and sum over the $np$ and $pp$
contributions as is done in obtaining the $NA$ effective interaction, and concentrate on the energy regime between 65 and 155~MeV laboratory kinetic energies. As guide to the eye we also include the corresponding quantity obtained from the high-precision Cd-Bonn potential~\cite{Machleidt:2000ge}, which fits $NN$ data with $\chi^2\approx 1$ up to 300 MeV. We focus on momentum transfers below 1.6~fm$^{-1}$, since the forward direction of the amplitude $A$ can be directly related to the differential cross sections
for $NA$ scattering at low momentum transfer (small angles)~\cite{Baker:2021iqy}. 
It is interesting to notice that while
at the smallest  energy (65 MeV) all three potential exhibit roughly the same central strength and agree with
the Cd-Bonn calculation, when moving to higher energies the NNLO$_{\rm{opt}}$ becomes considerably weaker,
specifically when extrapolating beyond the energy range included when fitting its low-energy constants. 
We should thus expect that those differences become visible in  $NA$ scattering observables.   

The Wolfenstein amplitude $C$ characterizes the spin-orbit force of the $NN$ interaction and is depicted in
Fig.~\ref{fig2} for three different $NN$ laboratory energies as function of the momentum transfer. In general,
this amplitude is much smaller in magnitude compared to $A$. We also see that here even at the lowest energy
the NNLO$_{\rm{opt}}$ is slightly larger compared to the amplitudes extracted from the other two interactions. 
This trend increases as the energy increases, indicating that the $NN$ spin-orbit force is slightly
stronger in the NNLO$_{\rm{opt}}$ chiral interaction. 
The amplitudes $C$ is the main contribution to the spin-orbit part of the $NA$
effective interaction, Eq.~(\ref{eq:1}), and thus it may be expected that its effect may be visible in 
the $NA$ spin observables. For a more in depth discussion we refer to Section~\ref{sec:scresults}.

Tensor force contributions of the $NN$ force are summed up in the Wolfenstein amplitudes $M$, $G$, and $H$. 
For $NA$ scattering from a $0^+$ target, only the amplitude $M$ contributes to the spin-orbit part of the
effective interaction. All three chiral interactions considered in this work give identical $M$ amplitudes 
in the energy range considered and agree perfectly well with the amplitude extracted from the Cd-Bonn
potential. A possible explanation may be that the tensor force contributions are determined quite well 
when fitting to deuteron properties. All $NN$ chiral interactions considered here are fitted to deuteron
properties. In addition, the effect of the tensor force contribution to $NA$ spin-observables is quite small
in case of scattering from $0^+$ targets, as was shown in Ref.~\cite{Burrows:2020qvu}.


\section{Structure Observables}
\label{sec:stresults}

Before discussing our results for nucleon-nucleus scattering, let us
first consider the results for the ground states of the three nuclei
under consideration, $^4$He, $^{12}$C, and $^{16}$O.  We obtain the
ground state energies and wavefunctions of these nuclei by employing
the NCSM~\cite{Barrett:2013nh}.  In this approach the $A$-body
wavefunctions are expanded in Slater determinants of $A$
single-particle wavefunctions which turns the $A$-body Schr\"odinger
equation into an eigenvalue problem for the expansion coefficients,
with the lowest eigenvalue being the ground state energy.  For any
finite basis expansion, the obtained eigenvalue, $E$, gives a strict
upper bound for the energy in the complete, but infinite-dimensional
basis, and the corresponding eigenvector gives an approximation to the
$A$-body wavefunction.  As one increases the basis size, the obtained
eigenvalues $E$ approach the exact eigenvalues for a given
Hamiltonian.

Here we follow the standard practice to use a harmonic oscillator (HO)
basis for the single-particle wavefunctions; for the many-body
truncation we use a truncation on the total number of HO quanta, that
is, a truncation on $\sum_i(2n_i+l_i)$ over all $A$ nucleons, with $n$
and $l$ the radial and orbital quantum numbers.  The actual truncation
parameter $N_{\max}$ is defined as the total number of HO quanta above
the minimal configuration for the ground state satisfying the Pauli
principle.  One of the reasons for using a HO basis, in combination
with a many-body truncation on the total number of HO quanta, is that
this leads to an exact factorization of the $A$-body wavefunction into
a center-of-mass wavefunction and a relative
wavefunction~\cite{Lipkin:1958zz,Gloeckner:1974sst}.  We use the
ground state wavefunction obtained in the NCSM to evaluate the
nonlocal one-body density in single-particle coordinates, from which
we subsequently obtain the translationally-invariant nonlocal one-body
density~\cite{Burrows:2017wqn} that is used as input to the $NA$
scattering calculation~\cite{Burrows:2020qvu}.

All three nuclei under consideration here have $0^+$ ground states;
both $^4$He and $^{16}$O are closed-shell nuclei, whereas $^{12}$C has
a deformed ground state.  For $^4$He and $^{16}$O, the most important
ground state observables are the energy and the charge radius; while
for $^{12}$C the quadrupole moment of the first excited state, which
is a rotational excitation of the ground state with $J=2$, also gives
information about the deformation.  The experimental charge radius is
related to the point-proton structure radius via
\begin{equation}
  r_{\rm str}^2 = r_{\rm charge}^2 - \left( R_p^2 + \frac{N}{Z} R_n^2 + \frac{3}{4 m_p^2} \right) \,,
  \label{Eq:protonradius}
\end{equation}
with $R_p^2$ and $R_n^2$ the proton and neutron mean-square charge
radii (note that $R_n^2$ is negative) , and $m_p$ is the proton mass.

Of course, the NCSM calculations depend on the truncation parameter,
$N_{\max}$, and the HO parameter, $\hbar\omega$; only in the limit of
$N_{\max}\to\infty$ do the physical observables become independent of
$\hbar\omega$.  For $^4$He, we can perform our numerical calculations
at sufficiently large $N_{\max}$ so that our wavefunctions are almost
converged, but for $^{12}$C (see Fig.~\ref{Fig:Egs_radius_12C}) and
$^{16}$O we are limited to $N_{\max}=10$ on current computational
resources.  This is sufficiently large to perform an extrapolation of
the obtained ground state energies to the complete basis, see
Table~\ref{Tab:Egs_radius}, but radii, and more general, densities, tend
to converge significantly slower, as can be seen in
Fig.~\ref{Fig:Egs_radius_12C} for $^{12}$C; the $\hbar\omega$ and
$N_{\max}$ dependence for $^{16}$O is similar.  We therefore
give in Table~\ref{Tab:Egs_radius} our result for the point-proton 
structure radii at the largest $N_{\max}$ over the range
$16$~MeV~$ < \hbar\omega < 24$~MeV
($15$~MeV~$ < \hbar\omega < 25$~MeV for Daejeon16)
which is a compromise of a common range for all three potentials in
which both the energy converges reasonably well and the radius appears
to converge.  This is also the range in $\hbar\omega$ that we use for
our calculations of scattering observables in
Sec.~\ref{sec:scresults}.

\begin{table}[tb]
  \begin{tabular}{l|l|l|l}
     & \multicolumn{1}{c|}{$^4$He} & \multicolumn{1}{c|}{$^{12}$C} & \multicolumn{1}{c}{$^{16}$O}    \\ \hline
     & \multicolumn{3}{c}{Ground state energy (MeV)}        \\ \hline
     Daejeon16      & $-28.372(1)$& $-92.9(0.1)$  & $-131.3(0.3)$ \\
     NNLO$_{\rm opt}$&$-27.596(1)$& $-88.4(1.3)$  & $-133.(3.)$   \\
     EKM N2LO       & $-28.11(1)$ & $-93.8(3.3)$  & $-149.(5.)$   \\
     expt.          & $-28.296$   & $-92.16$      & $-127.62$     \\ \hline
     & \multicolumn{3}{c}{RMS point-proton structure radius (fm)} \\ \hline
     Daejeon16      & $1.514 \sim 1.513$ & $2.30 \sim 2.27$ & $2.40 \sim 2.32$ \\
     NNLO$_{\rm opt}$&$1.440 \sim 1.436$ & $2.30 \sim 2.14$ & $2.35 \sim 2.17$ \\
     EKM N2LO       & $1.450 \sim 1.416$ & $2.29 \sim 2.05$ & $2.28 \sim 2.05$ \\
     expt.          &\multicolumn{1}{c|}{$1.484(6)$}&
                      \multicolumn{1}{c|}{$2.341(5)$}&\multicolumn{1}{c}{$2.58(1)$}\\ 
  \end{tabular}
  \caption{Ground state energy (top, extrapolated to the complete basis)
      and point-proton radius (bottom)
      obtained with the Daejeon16 potential~\cite{Shirokov:2016ead},
      the NNLO$_{\rm opt}$~\cite{Ekstrom13} chiral potential,
      and the EKM chiral potential at order N2LO~\cite{Epelbaum:2014sza,Epelbaum:2014efa},
      as well as the experimental values.
      \label{Tab:Egs_radius}
      }
\end{table}        

Numerical convergence is significantly better with Daejeon16 than
with the other two potentials, as one would expect: although Daejeon16
is based on a chiral EFT potential, like the other two potentials,
Daejeon16 is specifically designed to have improved numerical
convergence of NCSM calculations by applying an SRG evolution on the
initial chiral EFT $NN$ potential~\cite{Shirokov:2016ead}.  On the other
hand, the EKM N2LO potential has the slowest convergence, both for the
energies and for the radii.

Table~\ref{Tab:Egs_radius} clearly shows that Daejeon16 also gives the
best agreement with the experimental values for these three nuclei --
which is not surprising since the ground state energies of these
nuclei (as well as several others) were included in the fitting of
Daejeon16~\cite{Shirokov:2016ead}.  Interestingly, NNLO$_{\rm opt}$
leads to noticeable underbinding of both $^4$He and $^{12}$C, while all
three interactions overbind $^{16}$O, though by different amounts.

The obtained point-proton structure radii are in reasonable agreement
with experiment for $^4$He and slightly too small for $^{12}$C, but
significantly too small for $^{16}$O, although it is hard to quantify
this discrepancy due to the slow convergence of radii in a HO basis.
The structure radii being too small for $^{16}$O is correlated with
the overbinding of $^{16}$O; indeed, the deviation between experiment
and calculation in both the ground state energy and the structure
radius is largest with the EKM N2LO potential and smallest with
Daejeon16.

Although neither the radius of the $^{12}$C ground state, nor the
quadrupole moment of the $J=2$ rotational excitation, are very well
converged, as shown in Ref.~\cite{Caprio:2022mkg} for a subset of the
same NN interactions, the dimensionless ratio $Q/r_{\rm str}^2$ is
much better converged, and for all three interactions here it is in
agreement with the experimental value, given the large experimental
uncertainties.  The obtained excitation energy is also in reasonable
agreement with experiment, to within the estimated uncertainties in
our calculations.

Before moving to the scattering applications, it is also illustrative
to look at the local densities of these three nuclei, with these three
different $NN$ interactions, as well as their charge form factors.
In Fig.~\ref{Fig:localdens} we show the local point-proton densities,
normalized so they integrate to the total number of protons.
The colored bands in this and subsequent figures correspond to the
same range in the HO parameter as used for the radii listed in
Table~\ref{Tab:Egs_radius}, $16$~MeV~$ < \hbar\omega < 24$~MeV
($15$~MeV~$ < \hbar\omega < 25$~MeV for Daejeon16), with the dashed
and solid lines our results at $\hbar\omega = 20$~MeV.  We clearly see
that the local density is spread out furthest with Daejeon16, whereas
the EKM N2LO potential leads to the most compact densities, as is also
evident from the differences in the structure radii in
Table~\ref{Tab:Egs_radius}; furthermore we see that the numerical
convergence is significantly better with Daejeon16 than with the other
two potentials, again, as expected.

Finally, we can calculate the electric charge form factor from the
Fourier transform of the one-body densities, ignoring contribution
from two-body currents.  For point-nucleons, the charge form factor
would simply be the Fourier transform of the proton density.  In order
to compare with the experimental form factors however, we do need to
incorporate the fact that the protons and neutrons are not
point-particles.  For the longitudinal form factor we have
\begin{eqnarray}
  F(q) &=& \frac{1}{Z}
  \frac{G_E^p(Q_{\rm el}^2) \, F_p(q)
      + G_E^n(Q_{\rm el}^2) \, F_n(q)}{\sqrt{1 + Q_{\rm el}^2/(4 m_N^2)}} \,,
\label{Eq:formfactor}
\end{eqnarray}
where $F_{p,n}(q)$ are the Fourier transforms of the local
point-proton and -neutron densities, $G_E^{p,n}(Q^2)$ the proton
and neutron electric form factors, and
$Q_{\rm el}^2 = q^2 - \omega_{\rm el}^2$ the four-momentum squared,
with $\omega_{\rm el} = \sqrt{q^2 + m_A^2} - m_A$ the energy transfer
corresponding to the elastic peak and $m_A$ the mass of the nucleus.
Here we take the parameterization of \cite{Kelly:2004hm} for the proton
and neutron form factors $G_E^{p,n}(Q^2)$.
Note that the charge radius in Eq.~(\ref{Eq:protonradius})
is related to the form factor via
\begin{equation}
  r_{\rm charge}^2 = -6 \frac{{\rm d}F(q^2)}{{\rm d}q^2} \Big|_{q=0} \,,
  \label{Eq:radiusffderivative}
\end{equation}
and similarly, the proton and neutron radii $R_{p,n}^2$ to the
derivatives of $G_E^{p,n}$, and the point-proton structure radius to
derivative of $F_{p}$; indeed, Eq.~(\ref{Eq:protonradius}) follows
directly from Eq.~(\ref{Eq:formfactor}) by taking the derivative with
respect to $q^2$, followed by setting $q = 0$.

In Figs.~\ref{Fig:ffc_4He} and \ref{Fig:ffc_12C16O} we show our
results for the charge form factors, compared to experimental data.
Interestingly, although the obtained form factors for $^4$He agrees
with experiment at small momenta (as could be expected based on the
agreement with experiment for the point-proton structure radius),
it disagrees for $q > 2~{\rm fm}^{-1}$, for all three interactions.
In particular, the first zero-crossing happens between 4.4 and
4.8~fm$^{-1}$, whereas the experimental data indicate a zero-crossing
at or just above 3~fm$^{-1}$.  It is unclear what the origin is of the
significant deviation with experiment starting above about
2.5~fm$^{-1}$ -- it could be caused by the lack of 3-body forces, the
lack of consistent 2-body currents, or simply an indication that the
chiral expansion breaks down for larger momenta.  Note that, quantum
Monte Carlo calculations, both with a phenomenological potential (AV18
plus UIX 3-body forces) and with local N$^2$LO chiral EFT
interactions, including 3-body forces give much better agreement with
experiment than our current calculations~\cite{Lynn:2017fxg}.

In contrast the calculated form factors for $^{12}$C and $^{16}$O
agree qualitatively with experiment up through the first zero-crossing,
with the first zero-crossing at only slightly larger momenta $q$ than
experimentally observed; this is likely related to the radii being
slightly too small.  Furthermore, the relative location of the first
zero-crossings in $^{12}$C and $^{16}$O does agree with the relative
differences in the radii, see Table~\ref{Tab:Egs_radius}.
Qualitatively similar results were obtained for $^{12}$C and $^{16}$O with
quantum Monte Carlo calculations, both with a phenomenological
potential (AV18 plus 3-body forces) and with local N$^2$LO chiral EFT
interactions, including 3-body forces~\cite{Lonardoni:2018nob}.

We should point out that the first minima in the charge form factors
for $^{12}$C and $^{16}$O occur at momentum transfers smaller than
2~fm$^{-1}$, that is, at much smaller momentum transfers than for
$^4$He.  Thus it is an open question if the failure in describing the
charge form factor of $^4$He results from inherent deficiencies in the
high momentum behavior of NCSM one-body densities, or from e.g. the
lack of 2-body current operators.

Lastly, note that our numerical convergence is best with Daejeon16, as
expected; but more interesting is that with Daejeon16 the magnitude of
the form factor drops off (much) more rapidly than with the other
potentials, not only for $^4$He, but also for $^{12}$C and $^{16}$O.
This is likely a direct consequence of the SRG evolution, which
suppresses the high-momentum components.

\section{Proton-Nucleus Scattering Observables}
\label{sec:scresults}

In this section we present calculations of observables for elastic scattering from the closed shell nuclei
$^4$He and $^{16}$O and the open-shell nucleus $^{12}$C 
based on the three different $NN$ interactions discussed in Section~\ref{sec:formal}. The
effective interactions are computed based on Eq.~(\ref{eq:1}), which indicates that the $NN$ interactions are
treated on the same footing in the structure as well as reaction part. For the laboratory kinetic energies we
choose as lowest energy 65~MeV being at the lower limit where the leading order in the spectator expansion
may be considered reliable, around 100~MeV where all interactions still describe the $NN$ Wolfenstein amplitudes
reasonably well, and a higher energy, around 160~MeV, where differences in the description of the $NN$
Wolfenstein amplitudes become already quite visible. The choice of the specific projectile energies for
each target nucleus is based on the availability of experimental data.

The differential cross sections divided by the Rutherford cross section are shown in Fig.~\ref{cs4He_log} for
proton scattering from $^4$He, in Fig.~\ref{cs12C_log} from $^{12}$C, and in Fig.~\ref{cs16O_log} from $^{16}$O.
In all figures the lines represent  the value for $\hbar\omega$=20~MeV, while the bands indicate the
variation in $\hbar\omega$ between 16 and 24 for the NNLO$_{\rm opt}$ and EKM interactions and between
15 and 25 for the Daejeon16 interaction. This dependence on $\hbar\omega$ results from the calculation
of the scalar and spin-projected one-body densities entering the $NA$ effective interaction in leading
order in the spectator expansion, and should become insignificant when $N_{\rm max}$ is large. 
This is quite well illustrated by contrasting the calculations of $^4$He,
which employ $N_{\rm max}=18$ to those of $^{12}$C and $^{16}$O for which only $N_{\rm max}=10$ is
computationally feasible. 

\begin{figure}
\begin{center}
\includegraphics[width=0.65\textwidth]{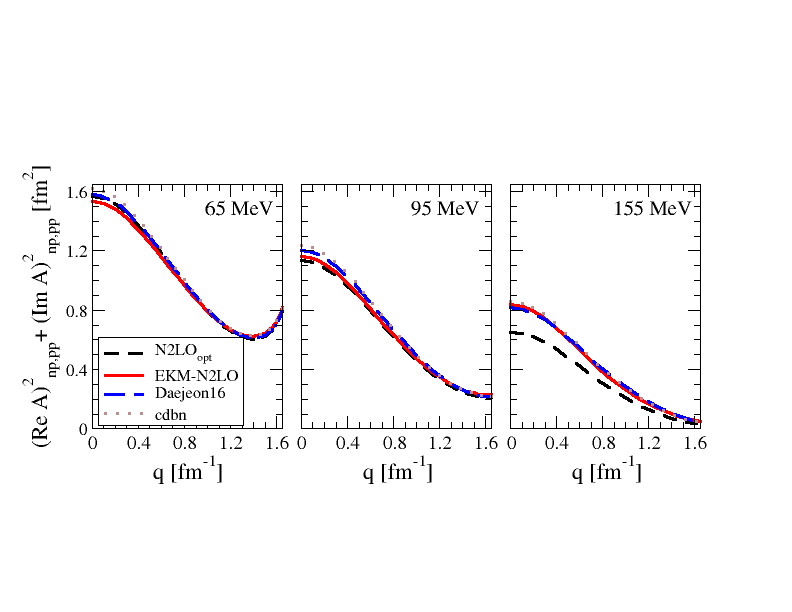}
\caption{The sum of the squares of the real and imaginary part of the Wolfenstein amplitude $A$ for $pp$ and $np$ scattering
as function of the momentum transfer $q$ at laboratory energies 65, 95, and 155 MeV calculated from the NNLO$_{\rm{opt}}$~\cite{Ekstrom13} chiral potential (short-dashed),
the EKM chiral potential at order N2LO~\cite{Epelbaum:2014sza,Epelbaum:2014efa} (solid), and the Daejeon16 potential~\cite{Shirokov:2016ead} (dashed).
As guide to the eye the dotted line represents the same quantity calculated from the Cd-Bonn~\cite{Machleidt:2000ge} high precision $NN$ potential.
} 
\label{fig1}
\end{center}
\end{figure}

\begin{figure}[hb]
\begin{center}
\includegraphics[width=0.65\textwidth]{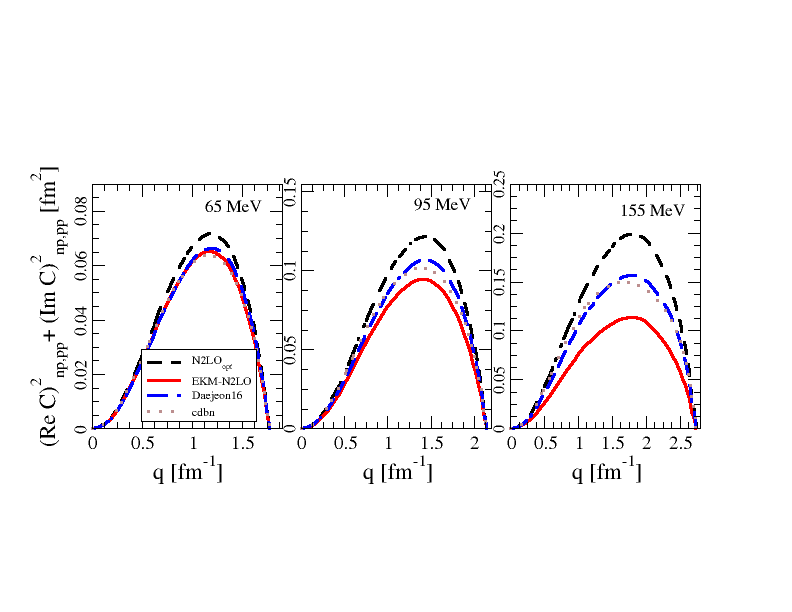}
\caption{The sum of the squares of the real and imaginary part of the Wolfenstein amplitude $C$ for $pp$
and $np$ scattering as function of the momentum transfer $q$ at laboratory energies 65, 95, and 155 MeV calculated from the
NNLO$_{\rm{opt}}$~\cite{Ekstrom13} chiral potential (short-dashed),
the EKM chiral potential at order N2LO~\cite{Epelbaum:2014sza,Epelbaum:2014efa} (solid), and the
Daejeon16 potential~\cite{Shirokov:2016ead} (dashed).
As guide to the eye the dotted line represents the same quantity calculated from the
Cd-Bonn~\cite{Machleidt:2000ge} high precision $NN$ potential.
}
\label{fig2}
\end{center}
\end{figure}

\begin{figure}[tb]
  \begin{center}
    \includegraphics[width=0.7\textwidth]{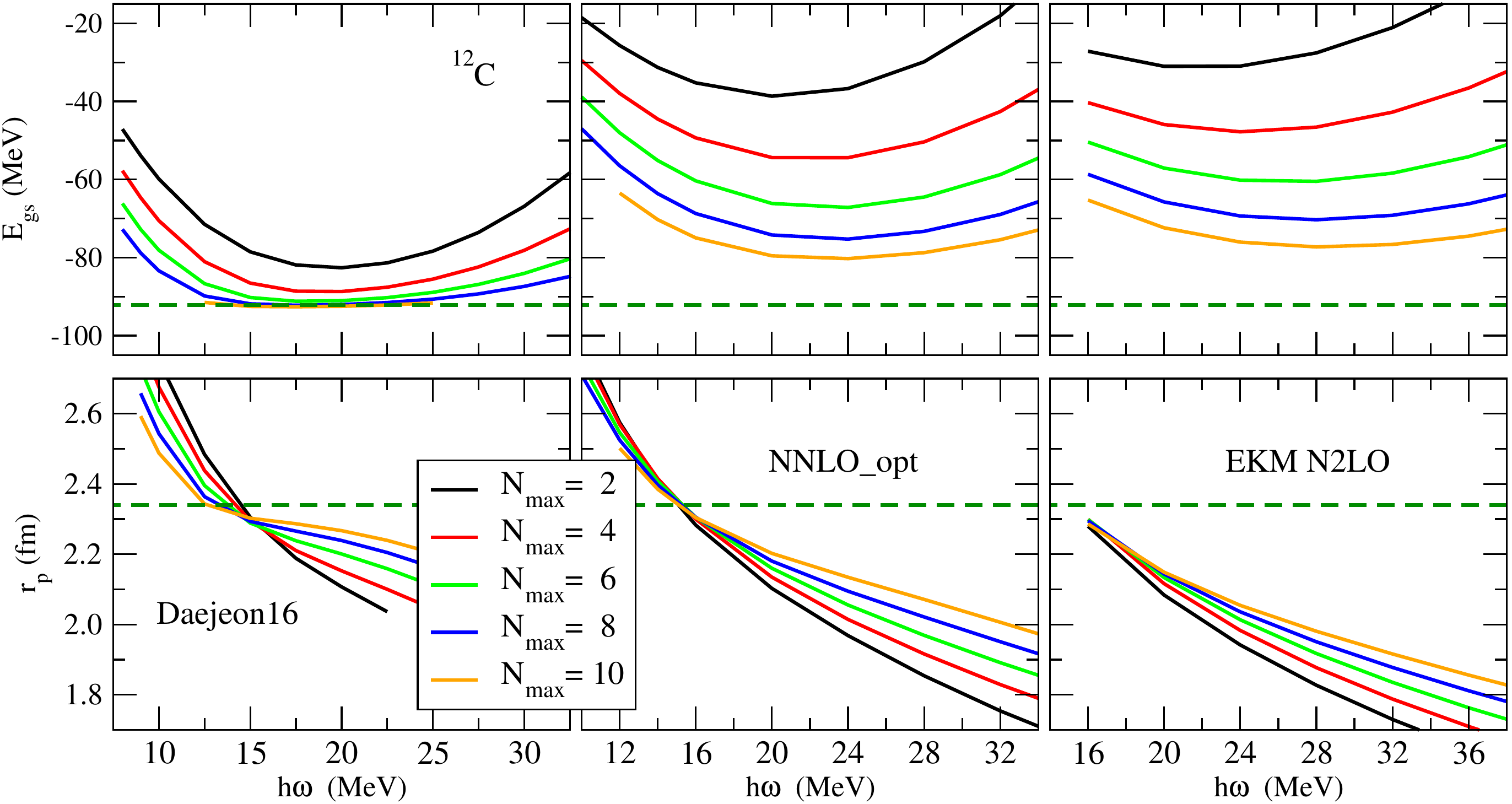}
    \caption{Ground state energy (top) and point-proton radius (bottom) of $^{12}$C
      obtained with the Daejeon16 potential~\cite{Shirokov:2016ead} (left),
      the NNLO$_{\rm opt}$~\cite{Ekstrom13} chiral potential (center),
      and the EKM chiral potential at order N2LO~\cite{Epelbaum:2014sza,Epelbaum:2014efa} (right).
      The dashed horizontal lines are the experimental values.
      \label{Fig:Egs_radius_12C}
      }
  \end{center}
\end{figure}

\begin{figure}
  \begin{center}
    \includegraphics[width=0.4\textwidth]{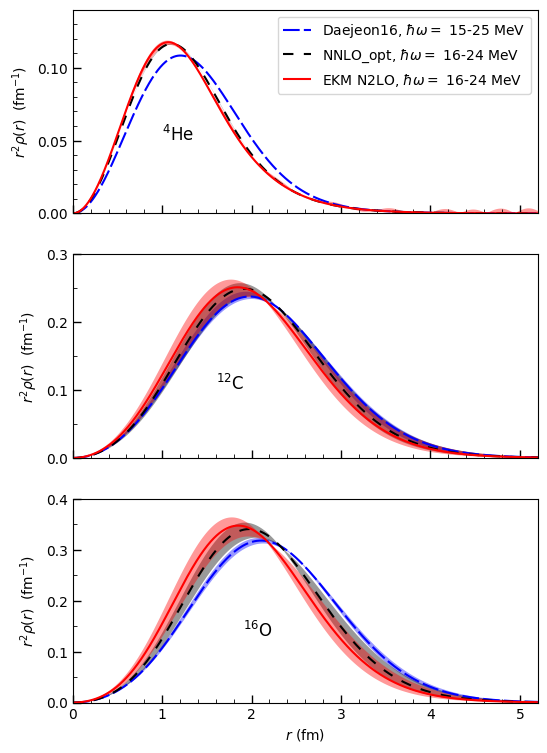}
    \caption{The local proton density, multiplied by $r^2$, of $^4$He (top),
      $^{12}$C (middle), and $^{16}$O (bottom), obtained with     
      the NNLO$_{\rm opt}$~\cite{Ekstrom13} chiral potential (short-dashed),
      the EKM chiral potential at order N2LO~\cite{Epelbaum:2014sza,Epelbaum:2014efa} (solid),
      and the Daejeon16 potential~\cite{Shirokov:2016ead} (dashed).
      The calculations for $^4$He are done at $N_{\rm max}=18$,
      those for $^{12}$C and $^{16}$O at $N_{\rm max}=10$, all with $\hbar\omega=20$ MeV. 
      The variation of $\hbar\omega$ from 16~MeV to 24~MeV for the NNLO$_{\rm opt}$ and EKM potentials,
      and from 15~MeV to 25~MeV for Daejeon16, are indicated by the shaded bands.
      \label{Fig:localdens}
      }
  \end{center}
\end{figure}

\begin{figure}
  \begin{center}
    \includegraphics[width=0.4\textwidth]{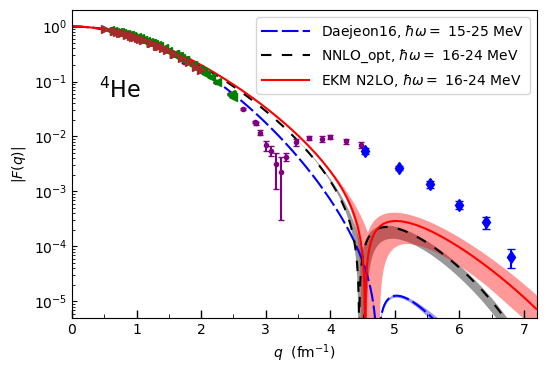}
    \caption{The calculated charge form factor of $^4$He, with experimental data
      from Refs.~\cite{Frosch:1967pz,Mccarthy:1977vd,Arnold:1978qs,Ottermann:1985km}.
      Colors, symbols, and calculational details are the same as in Fig.~\ref{Fig:localdens}.
      \label{Fig:ffc_4He}
    }
  \end{center}
\end{figure}

\begin{figure}
  \begin{center}
    \includegraphics[width=0.4\textwidth]{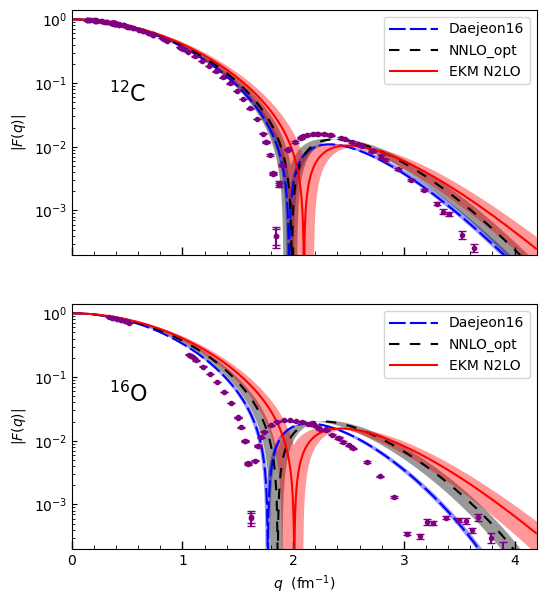}
    \caption{The charge form factor of $^{12}$C (top) and $^{16}$O (bottom),
      with a compilation of experimental data \cite{unpublished12C16OSick,private12C16OLonardoni}
      based on Refs.~\cite{schutz1975elastic,Sick:1970ma}.
      Colors, symbols, and calculational details are the same as in Fig.~\ref{Fig:localdens}.
      \label{Fig:ffc_12C16O}
    }
  \end{center}
\end{figure}

\begin{figure}
\begin{center}
\includegraphics[width=0.4\textwidth]{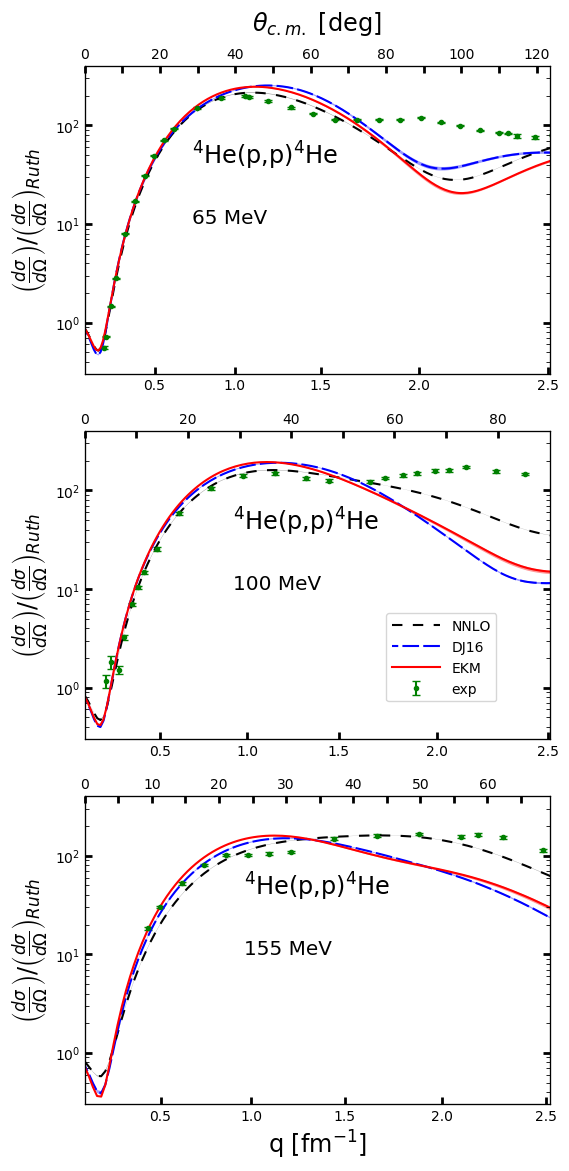}
\caption{The differential cross section divided by the Rutherford cross section for proton scattering
from $^4$He at 65 MeV (top), 100~MeV (middle), and 155~MeV (bottom) calculated using the NNLO$_{\rm{opt}}$~\cite{Ekstrom13}
chiral potential (short-dashed), the EKM chiral potential at order
N2LO~\cite{Epelbaum:2014sza,Epelbaum:2014efa} (solid), and the
Daejeon16 potential~\cite{Shirokov:2016ead} (dashed). All calculations use $N_{\rm max}=18$ (and 20 for 
the EKM potential) and $\hbar\omega=20$ MeV. 
The variation of $\hbar\omega$ from 16~MeV to 24~MeV for the NNLO$_{\rm{opt}}$ and EKM  and
from 15~MeV to 25~MeV for the Daejeon16 structure calculations are indicated by the bands.
The experimental data at 65~MeV are from Refs.~\cite{Imai:1979ihs}, at 
100~MeV from \cite{Goldstein:1970dg} and at 155~MeV from \cite{Comparat:1975bm}.
}
\label{cs4He_log}
\end{center}
\end{figure}

\begin{figure}
\begin{center}
\includegraphics[width=0.4\textwidth]{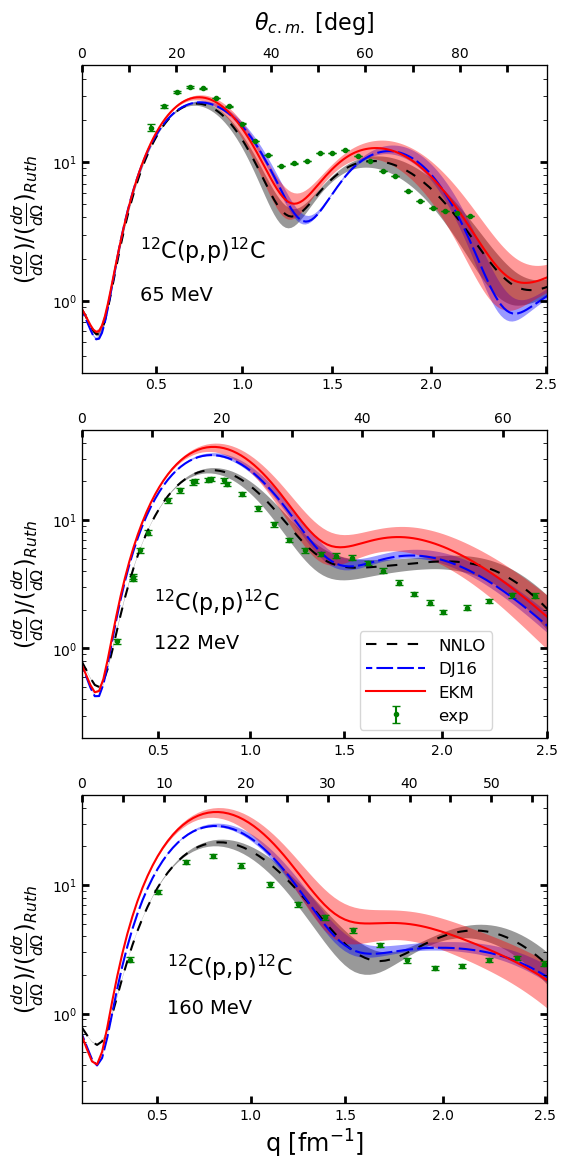}
\caption{The differential cross section divided by the Rutherford cross section for proton scattering 
from $^{12}$C at 65 MeV (top), 122~MeV (middle), and 160~MeV (bottom) calculated using the
NNLO$_{\rm{opt}}$~\cite{Ekstrom13}
chiral potential (short-dashed), the EKM chiral potential at order
N2LO~\cite{Epelbaum:2014sza,Epelbaum:2014efa} (solid), and the
Daejeon16 potential~\cite{Shirokov:2016ead} (dashed). All calculations use $N_{\rm max}=10$. The lines 
indicate the value of $\hbar\omega=20$ MeV, while the bands characterize the variation of $\hbar\omega$ 
 from 16~MeV to 24~MeV for the NNLO$_{\rm{opt}}$ and EKM,  and
from 15~MeV to 25~MeV for the Daejeon16 structure calculations. 
The experimental data at 65~MeV are from Refs. ~\cite{Ieiri1987:sns}  
and \cite{Kato:1985koh}, at 122 and 160~MeV from \cite{Meyer:1983kd}.
}
\label{cs12C_log}
\end{center}
\end{figure}

\begin{figure}
\begin{center}
\includegraphics[width=0.4\textwidth]{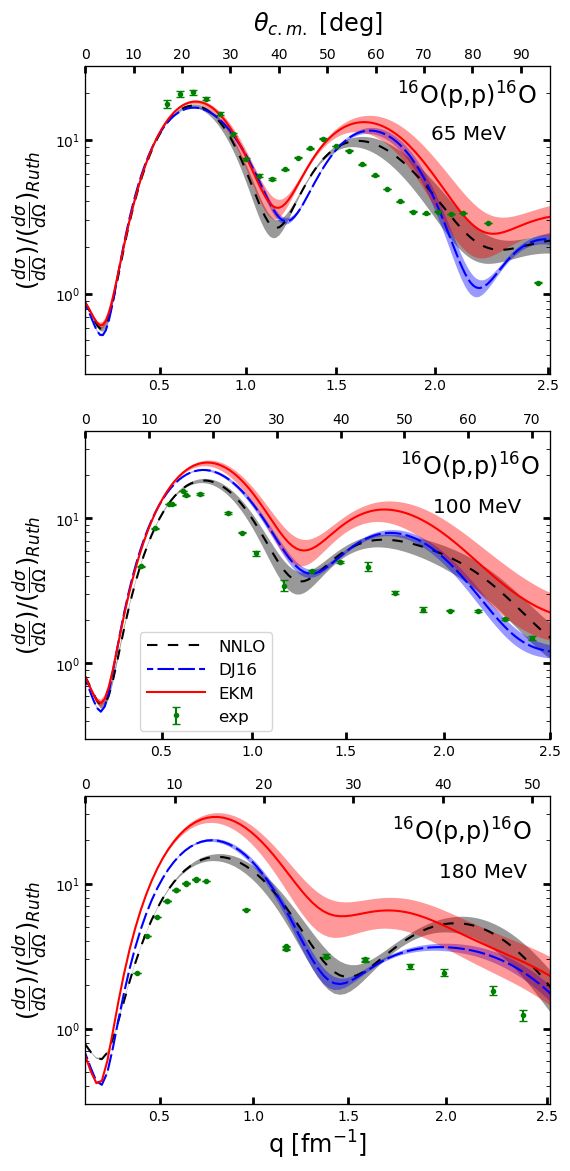}
\caption{The differential cross section divided by the Rutherford cross section for proton scattering
from $^{16}$O at 65 MeV (top), 100~MeV (middle), and 180~MeV (bottom) calculated using the
NNLO$_{\rm{opt}}$~\cite{Ekstrom13}
chiral potential (short-dashed), the EKM chiral potential at order
N2LO~\cite{Epelbaum:2014sza,Epelbaum:2014efa} (solid), and the
Daejeon16 potential~\cite{Shirokov:2016ead} (dashed). All calculations use $N_{\rm max}=10$. The lines
indicate the value of $\hbar\omega=20$ MeV. The bands characterize the variation of $\hbar\omega$ 
 from 16~MeV to 24~MeV for the NNLO$_{\rm{opt}}$ and EKM, and
from 15~MeV to 22.5~MeV for the Daejeon16 structure calculations.
The experimental data at 65~MeV are from Ref.~\cite{Sakaguchi:1979fpk}
at 100~MeV from~\cite{Seifert:1990um} and at 180~MeV from~\cite{Kelly:1990zza}. 
}
\label{cs16O_log}
\end{center}
\end{figure}

The leading order of the multiple scattering expansion is usually considered quite reliable at 100~MeV
projectile energy and higher, while when moving towards lower energies corrections may start to play a
role~\cite{Chinn:1994xz,Chinn:1995qn}. Considering the differential cross sections for all three nuclei
for energies around 100 MeV and higher and momentum transfers to about 1.5 fm$^{-1}$ we observe that all
calculations over-predict the experimental values, with the calculation based on the NNLO$_{\rm opt}$
interaction being closest to the data. 
This is especially noteworthy, since  the squares of the
summed $np$ and $pp$ Wolfenstein amplitudes $A$ shown at similar energies in Fig.~\ref{fig1}, 
which represent the central piece of the $NN$ interaction,
under-predict the corresponding $NN$ values represented by the Cd-Bonn values at for $NN$ energies
exceeding 100~MeV. 
As also shown in Fig.~\ref{fig1}, at 65 MeV all three $NN$ interactions essentially coincide even for very low
momentum transfers. This leads to very similar differential cross sections at this energy at least for
the $p$-shell nuclei $^{12}$C and $^{16}$O at the same energy, though here all calculations under-predict
the experimental values. 

The predictions of the differential cross section for $^4$He for the energies shown in
Fig.~\ref{cs4He_log}  do not describe the experiment beyond about $q \simeq 1$~fm$^{-1}$.
This may be related to the charge form factor of $^4$He, as calculated with these interactions,
being far from experiment in the first minimum as illustrated in Fig.~\ref{Fig:ffc_4He};
or it could be an indication of e.g. a deficiency of the leading-order spectator expansion
for the tightly-bound $s$-shell nucleus $^4$He at energies well below 200 MeV.
For both $p$-shell nuclei, $^{12}$C and $^{16}$O, the calculated first minimum of the charge form factor is
much closer to experiment, albeit at a slightly higher momentum transfer. This results in the maximum-minimum
structure within the differential cross sections for those nuclei also being shifted to slightly higher momentum
transfers. 
 
\begin{figure}
\begin{center}
\includegraphics[width=0.4\textwidth]{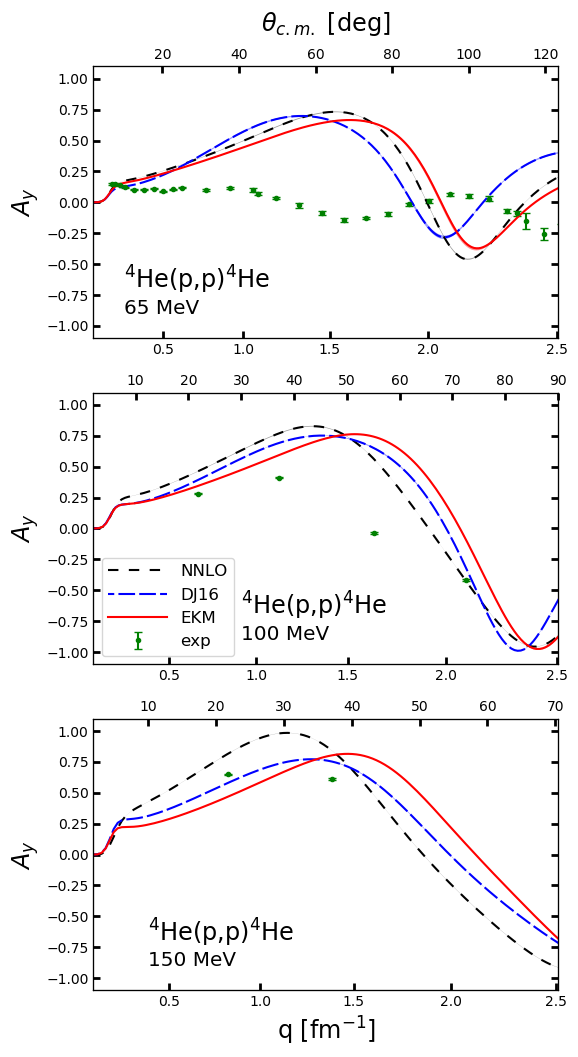}
\caption{The analyzing power $A_y$ for proton scattering
from $^{4}$He at 65 MeV (top), 100~MeV (middle), and 150~MeV (bottom) calculated using the
NNLO$_{\rm{opt}}$~\cite{Ekstrom13}
chiral potential (short-dashed), the EKM chiral potential at order
N2LO~\cite{Epelbaum:2014sza,Epelbaum:2014efa} (solid), and the
Daejeon16 potential~\cite{Shirokov:2016ead} (dashed). All calculations use $N_{\rm max}=18$ for NNLO$_{\rm{opt}}$ and Daejeon16 potentials, and $N_{\rm max}=20$ for the EKM potential. The lines
indicate the value of $\hbar\omega=20$ MeV. The bands characterize the variation of $\hbar\omega$ 
from 16~MeV to 24~MeV for the NNLO$_{\rm{opt}}$ and EKM,  and from 15~MeV to 25~MeV for the Daejeon16 structure
 calculations.
The experimental data at 65~MeV are from Ref.~\cite{Imai:1979ihs}
        at 100~MeV and 150~MeV from~\cite{Wesick:1986bt}.
}
\label{Ay4He}
\end{center}
\end{figure}

\begin{figure}
\begin{center}
\includegraphics[width=0.4\textwidth]{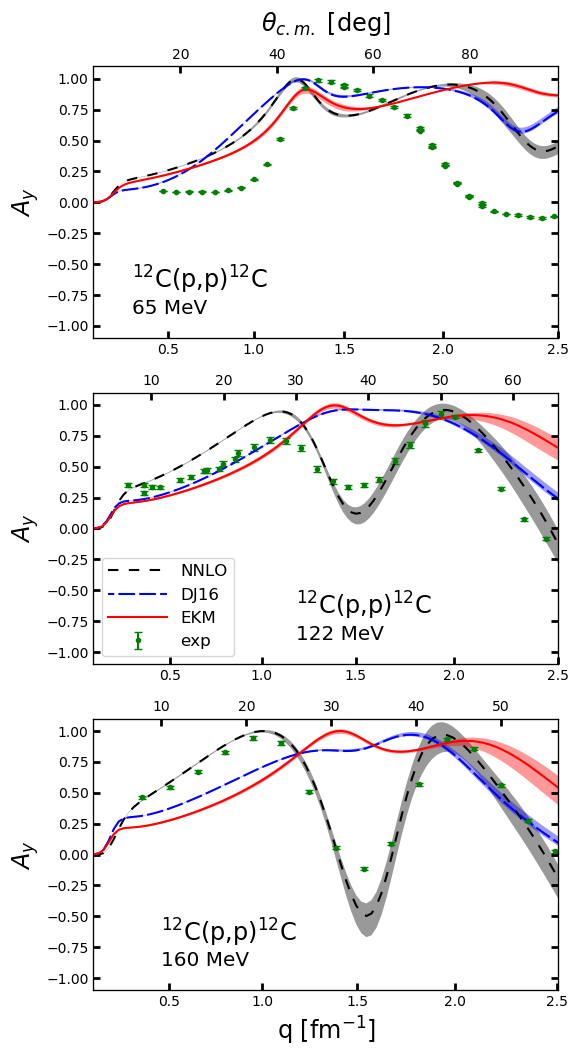}
\caption{The analyzing power $A_y$ for proton scattering
from $^{12}$C at 65 MeV (top), 122~MeV (middle), and 160~MeV (bottom) calculated using the
NNLO$_{\rm{opt}}$~\cite{Ekstrom13}
chiral potential (short-dashed), the EKM chiral potential at order
N2LO~\cite{Epelbaum:2014sza,Epelbaum:2014efa} (solid), and the
Daejeon16 potential~\cite{Shirokov:2016ead} (dashed). All calculations use $N_{\rm max}=10$. The lines
indicate the value of $\hbar\omega=20$ MeV. The bands characterize the variation of $\hbar\omega$
from 16~MeV to 24~MeV for the NNLO$_{\rm{opt}}$ and EKM, and
from 15~MeV to 25~MeV for the Daejeon16 structure calculations.
The experimental data at 65~MeV are from Ref.~\cite{Ieiri1987253}
        at 122~MeV and at 160~MeV from~\cite{Meyer:1983kd}.
}
\label{Ay12C}
\end{center}
\end{figure}

\begin{figure}
\begin{center}
\includegraphics[width=0.4\textwidth]{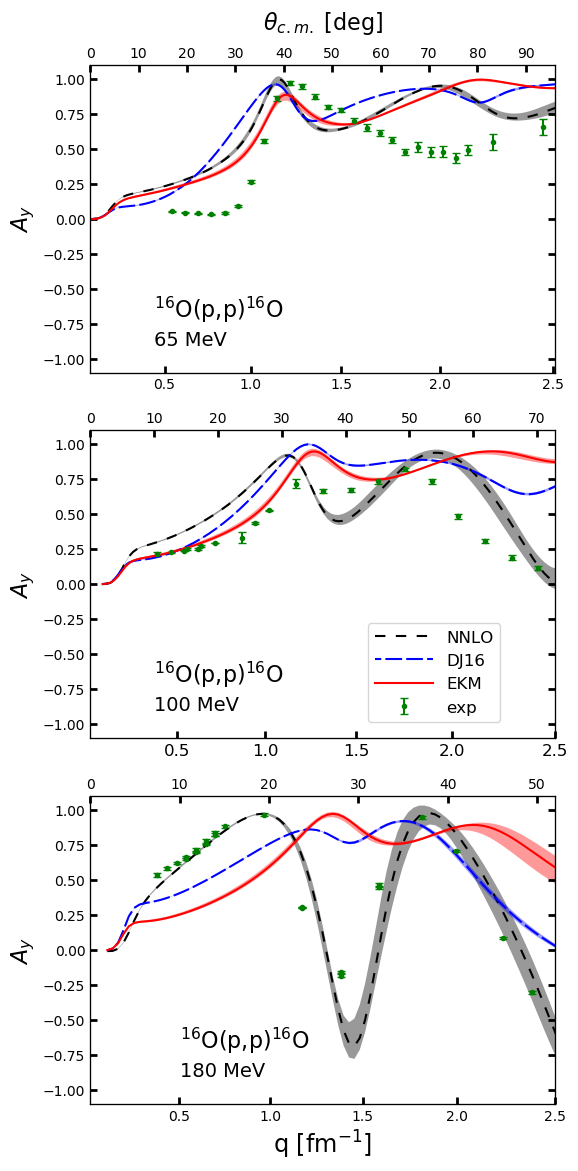} 
	\caption{ The analyzing power $A_y$ for proton scattering from $^{16}$O at 65 MeV (top), 100 MeV (middle), and 180~MeV (bottom) calculated using the
NNLO$_{\rm{opt}}$~\cite{Ekstrom13}
chiral potential (short-dashed), the EKM chiral potential at order
N2LO~\cite{Epelbaum:2014sza,Epelbaum:2014efa} (solid), and the
Daejeon16 potential~\cite{Shirokov:2016ead} (dashed). All calculations use $N_{\rm max}=10$. The lines
indicate the value of $\hbar\omega=20$ MeV. The bands characterize the variation of $\hbar\omega$
from 16~MeV to 24~MeV for the NNLO$_{\rm{opt}}$ and EKM,  and
from 15~MeV to 22.5~MeV for the Daejeon16 structure calculations.
The experimental data at 65~MeV are from Refs.~\cite{Sakaguchi:1979fpk}, at 100~MeV from 
\cite{Seifert:1990um}, and at 180~MeV from~\cite{Kelly:1990zza}.
}
\label{Ay16O}
\end{center}
\end{figure}

Next, we consider spin observables for the same nuclei at the same energies. There are two independent
spin observables that can be obtained for scattering of a spin-$\frac{1}{2}$ particle from a spin-0 target, namely the
analyzing power $A_y$ and the spin-rotation function $Q$. The analyzing power considers the projectile spin
normal to the scattering plane, while the the spin-rotation function refers to the change of the spin
direction in the scattering plane. To experimentally determine the latter, a double scattering measurement
must be performed, a reason why experimental information for the spin-rotation function is relatively scarce.
The analyzing powers for proton scattering from $^4$He are shown in Fig.~\ref{Ay4He}, from $^{12}$C in
Fig.~\ref{Ay12C}, and from $^{16}$O in Fig.~\ref{Ay16O}. The error bands represent different values of
$\hbar\omega$ in the calculation of the nonlocal one-body densities. The results for $^4$He use
$N_{\rm max}=18$, Fig.~\ref{cs4He_log} shows that those calculations are well converged. This is not the
case for $^{12}$C and $^{16}$O, where $N_{\rm max}=10$ is the highest value that can reasonably used in the
structure calculation. Nevertheless, for momentum transfers below 1.5~fm$^{-1}$ all calculations are
reasonably converged. For both $p$-shell nuclei, $^{12}$C and $^{16}$O, we observe a common behavior as
function of energy. At the lowest energy, 65~MeV, the experimental data for 
the analyzing power are consistent with zero up to about
1~fm$^{-1}$, and then quickly rise to their maximum value of one. The position of this first maximum is
captured by all three calculations reasonably well, while none of them reflects an almost zero 
analyzing power for
very small momentum transfers. For $^4$He at 65~MeV, the experimental values of the analyzing power is very
small over the entire range of momentum transfers, which is not captured by any of the calculations.
At 100~MeV the slow rise of $A_y$ at momentum transfers smaller than 1~fm$^{-1}$, is captured better by
calculations based on the EKM and Daejeon16 calculations, while the first dip is only reproduced by 
calculations based on the NNLO$_{\rm opt}$ interaction. At the highest energy shown, only the 
NNLO$_{\rm opt}$ interaction is able to describe the data, especially over the entire range of 
momentum transfers shown. 

To possibly gain some insight into the behavior of the analyzing power and its energy dependence, 
we need to remember, that the parameterization in terms of Wolfenstein amplitudes 
of Eq.~(\ref{eq:3}) can also be used to describe $NA$ scattering. The amplitude for the scattering of a spin-$\frac{1}{2}$
particle from a spin-0 particle is simpler than the one given in Eq.~(\ref{eq:3}) and only contains two
Wolfenstein amplitudes, namely $A(\bm{q},\bm{\mathcal{K}}_{NA})_{NA}$ and
$i C(\bm{q},\bm{\mathcal{K}}_{NA})_{NA} \: (\bm{\sigma^{(0)}}\cdot\hat{\bm{n}})$~\cite{RodbergThaler,RHLandau}.
In this case, there are no tensor amplitudes $M$, $G$, and $H$. For the calculation of observables,
only the on-shell pieces of the Wolfenstein amplitudes contribute, and the spin observables are given as
\begin{eqnarray}
A_y &=& \frac{2 \Re e \left[ A^*(\bm{q},\epsilon)_{NA} \: C(\bm{q},\epsilon)_{NA}\right]}
 {|A(\bm{q},\epsilon)_{NA}|^2 + |C(\bm{q},\epsilon)_{NA}|^2} \cr
Q &=& \frac{2 \Im m \left[ A(\bm{q},\epsilon)_{NA} \: C^*(\bm{q},\epsilon)_{NA}\right]}
{ |A(\bm{q},\epsilon)_{NA}|^2 + |C(\bm{q},\epsilon)_{NA}|^2},
\label{eq:ayq}
\end{eqnarray}
with ${\bf q}$ being the momentum transfer.
The unpolarized differential cross section is proportional to the denominator in the above
expressions. The expressions for $A_y$ as well as $Q$ show that those observables are given by products of
real and imaginary parts of the amplitude $A(\bm{q},\epsilon)_{NA}$ representing the
central part of the effective $NA$ interaction as well as the spin-orbit part $C(\bm{q},\epsilon)_{NA}$. 
They also show that they test different combinations of real and imaginary parts of the effective
interaction. Having this in mind, it should not be surprising, that a better description of the differential cross
section by the NNLO$_{\rm opt}$ interaction leads to a better description of $A_y$. 
This is consistent with
the findings in Ref.~\cite{Catacora-Rios:2020xgx}, which studied the overlap of parameter spaces when
fitting differential cross sections and $A_y$ data based on phenomenological optical potentials in a Bayesian
approach. They found that both sets of data ultimately lead to similar likelihood functions and thus similar
qualities of their fits.  Considering the $NN$ amplitudes for the spin-orbit contribution,
Fig.~\ref{fig2}, the NNLO$_{\rm opt}$ interaction has for all shown energies a stronger spin-orbit force
compared to the other two interactions. This together with a weaker central part seems to lead to a better
description of analyzing power for $^{12}$C and $^{16}$O especially at the higher energies. The prediction of
the analyzing power in $^4$He is far away from the experimental data. Referring to the observation and
discussion of the differential cross section for $^4$He, this should not be a surprise. In addition, $^4$He
is a tightly bound system predominantly in the $s$-state. The leading order in the
spectator expansion may not be sufficient for obtaining the effective $NA$ interaction.

\begin{figure}
\begin{center}
\includegraphics[width=0.4\textwidth]{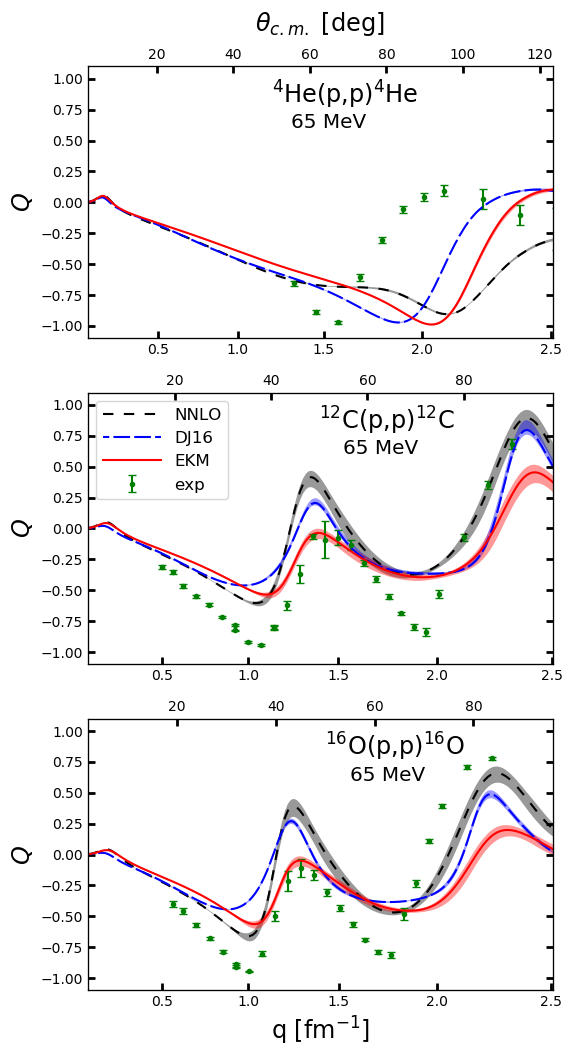}
\caption{The spin rotation function $Q$ for proton scattering from $^4$He (top), $^{12}$C (middle), and
$^{16}$O (bottom) at 65 MeV calculated using the
NNLO$_{\rm{opt}}$~\cite{Ekstrom13} chiral potential (short-dashed), the EKM chiral potential at order
N2LO~\cite{Epelbaum:2014sza,Epelbaum:2014efa} (solid), and the
Daejeon16 potential~\cite{Shirokov:2016ead} (dashed). The description of the lines and bands is the same as in
Figs.~\ref{Fig:ffc_12C16O} and~\ref{cs12C_log}. The experimental data are from Ref.~\cite{Sakaguchi1986}}

\label{Q_HeCO_65}
\end{center}
\end{figure}

Experimental information for the spin-rotation function $Q$ is considerable sparser compared to the ones for
$A_y$. For proton scattering at 65~MeV data exist for all nuclei we consider here and they are shown in
Fig.~\ref{Q_HeCO_65} together with the corresponding calculations. For both, $^{12}$C and $^{16}$O, the peak and
dip structure is captured by all three effective $NA$ interactions almost up to $q\approx 2$~fm$^{-1}$, 
though the magnitude is not. A study in Ref.~\cite{Chinn:1994xz} for heavier nuclei indicated 
that taking into account in an
approximate fashion pieces beyond the leading order in the spectator expansion may remedy this situation.
The light nucleus $^4$He again does not fit this scheme at all; here the first dip in the data is not captured by
the calculations at all, most likely because of the deficiencies already discussed earlier.


\section{Extraction of the local part of the Optical Potentials}
\label{sec:loptpots}

\begin{figure}
\begin{center}
\includegraphics[width=0.95\textwidth]{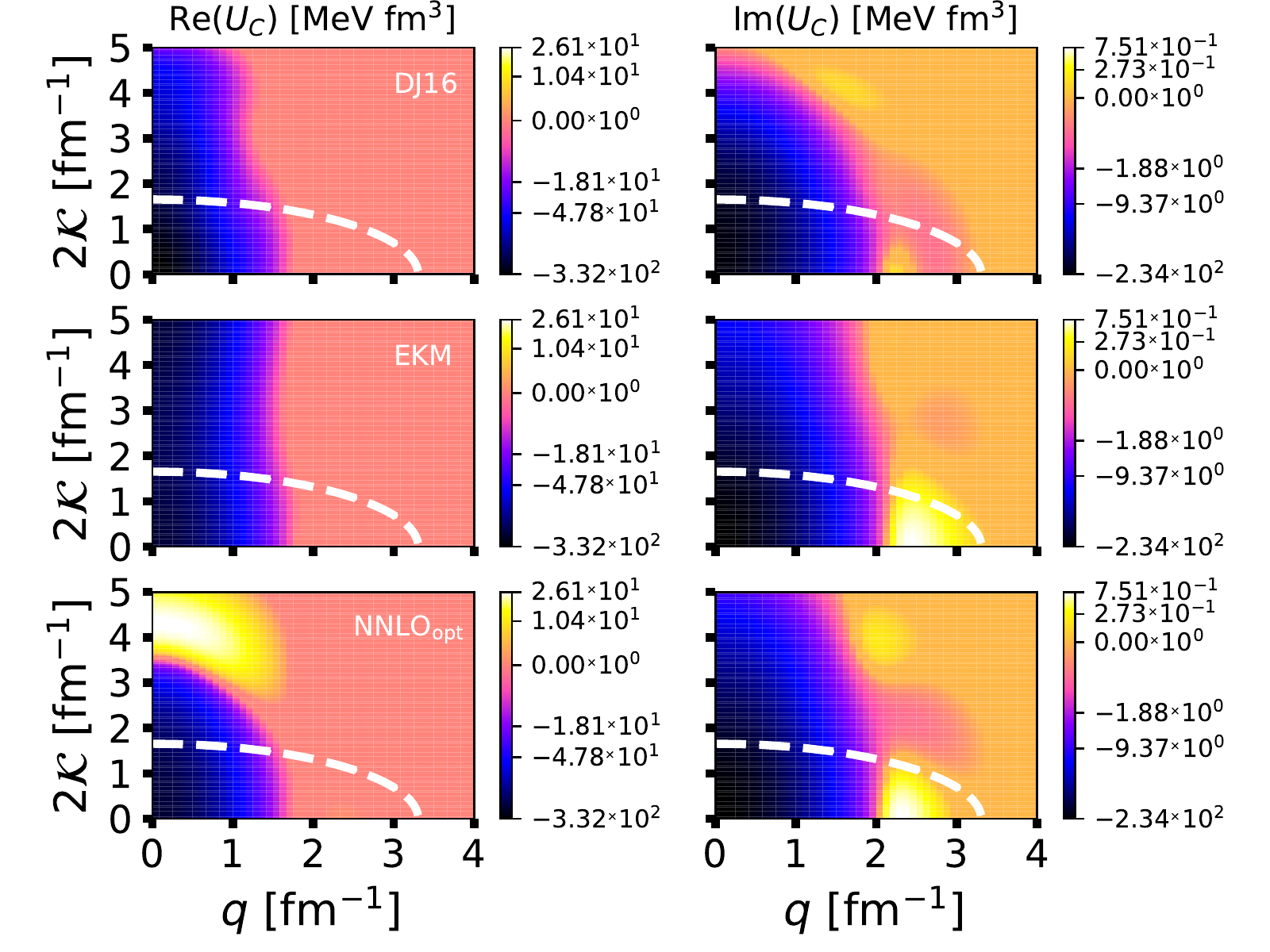}
\caption{The central part, $\widehat{U}_C (q,{\cal K},\theta_{q,{\cal K}})$, 
of the nonlocal effective interaction for 
proton scattering from $^{12}$C at 65~MeV
laboratory kinetic energy. The interaction is shown for the real (left panels) and imaginary (right panels)
parts of $U_C$ calculated from the Daejeon16 potential~\cite{Shirokov:2016ead} (upper row), 
the EKM chiral potential at order
N2LO~\cite{Epelbaum:2014sza,Epelbaum:2014efa} (middle row), and the NNLO$_{\rm{opt}}$~\cite{Ekstrom13} chiral
potential (bottom row).
The potentials are plotted for the angle $\theta_{q,{\cal K}}$=90$^o$ as functions of the
momentum transfer $q$ and the average momentum ${\cal K}$ in the NA c.m. frame.
The NCSM calculations use parameters $N_{\rm max}=10$ and $\hbar\omega=20$~MeV in all
cases. The dashed indicates the values of $U_C (q,{\cal K})$ for which the on-shell condition is fulfilled. 
}
\label{fig10}
\end{center}
\end{figure}

As seen in the previous Section, the {\it ab initio} effective $NA$ interactions derived from three different
chiral $NN$
interactions in leading order in the spectator expansion yield differences in $NA$ scattering observables
that are comparable or larger to uncertainties that are obtained by chiral effective field theory
uncertainties~\cite{Baker:2021iqy,Baker:2023uzx}. Though potentials are not observable, they can
provide a guide on differences in calculations thereof. Following Eq.~(\ref{eq:1}) the leading order
effective $NA$ interaction $\widehat{U}(\bm{q},\bm{\mathcal{K}}_{NA},\epsilon)$ is a scalar
function of two vector variables, $\bm{q}$ and $\bm{\mathcal{K}}_{NA}$, and an energy variable~$\epsilon$.  
The two momenta are explicitly given in Eq.~(\ref{eq:2}). Since the potential is considered
in the c.~m.~frame of the projectile-target $(A+1)$ system, we refer for simplicity to the average momentum
here as $\bm{\mathcal{K}}$ and omit the subscript ${\it NA}$, and also omit the dependence on
$\epsilon$. The potential has two pieces, a central and a spin-orbit part, which both are complex
functions.  
To illustrate the functional forms we show in Fig.~\ref{fig10} the central potential
$\widehat{U}_C (q,\mathcal{K},\theta_{q,\mathcal{K}})$ for 
proton scattering from $^{12}$C at 65~MeV laboratory kinetic energy 
obtained {\it ab initio} from the three different chiral $NN$ interactions we consider. 
The potentials are shown
for the fixed angle $\theta_{q,\mathcal{K}}=90^o$ between the 
vector variables $\bm{q}$ and $\bm{\mathcal{K}}$. We verified that potentials depend only weakly on the
angle between the two vectors, and thus the potential is almost separable in them as already pointed out in
Ref.~\cite{Arellano:2022tsi}.
This particular property is imprinted on the potential by the nonlocal
one-body density, as shown in Ref.~\cite{Burrows:2017wqn}.  

The potential surfaces for $\theta_{q,\mathcal{K}}=90^o$ are special, since for them the 
on-shell condition
\begin{equation}
q^2 + 4 \mathcal{K}^2 = 4 k_0^2 
\label{onshell}
\end{equation} 
is defined. Here $k_0$ is the on-shell momentum in the $NA$ c.~m.~frame.  The white dashed lines in
Fig.~\ref{fig10} indicate this on-shell condition. For this illustration we chose to show the
fully-off-shell central part of the effective interaction at 65~MeV, since at this energy the
chiral $NN$ amplitudes shown in Fig.~\ref{fig1}  given by the interactions are quite similar. However, 
the interactions lead to different off-shell behavior especially in the variable $\mathcal{K}$. 
As pointed out in Sec.~\ref{sec:formal}, the Daejeon16 interaction is based on the Idaho N3LO chiral
interaction, which is SRG evolved with a flow parameter $\lambda$=1.5~fm$^{-1}$. A characteristic of
an SRG evolution is that the interaction becomes softer while its on-shell characteristics is preserved. 
The effective interaction based on the Daejeon16 interactions falls off fastest as function of the
variable
$\mathcal{K}$, while both, the EKM and the NNLO$_{\rm opt}$ interactions exhibit stronger, albeit
different off-shell behavior in $\mathcal{K}$, especially in the real part of 
$\widehat{U}_C (q,\mathcal{K},\theta_{q,\mathcal{K}})$.

If we want a more quantitative view of the effective potentials, we should focus on a part which can be
connected to a local potential. It was shown in Ref.~\cite{Arellano:2022tsi} that the most important
piece of the potential contributing in the scattering calculation is the local part of the potential,
while the nonlocal part could be approximated with terms of a Gaussian non-locality. This means that the
far off-shell differences shown in Fig.~\ref{fig10} likely contribute little to elastic scattering observables. 
We want to pursue this line of thought further. With the help of the on-shell condition,
Eq.~(\ref{onshell}), we can obtain from the potential surfaces (central as well as spin-orbit)
$\widehat{U}(q,\mathcal{K},\theta_{q,\mathcal{K}}=90^o)$ a function $\widehat{U}(q)$ at a given
scattering energy,  which we then Fourier transform
to obtain functions $\widehat{U}_C(\zeta)$ and
$\widehat{U}_{SO}(\zeta)$. The radial variable $\zeta=\frac{1}{2}(r'+r)$ is the conjugate variable to the
momentum transfer $q$~\cite{Burrows:2017wqn}, if one would consider nonlocal potentials in coordinate
space as functions of $r$ and $r'$. To avoid confusion we refer to the local coordinate as $\zeta$.  

When calculating observables we first construct the Watson optical potential according to
Eq.~(\ref{eq:watson}). Since this procedure is not feasible in coordinate space we apply the well known
KMT factor (A-1)/A~\cite{KMT} to obtain local optical potentials 
$U_C(\zeta)$ and $U_{SO}(\zeta)$. The KMT approximation represents scattering observables obtained from
a  Watson
optical potential very well for momentum transfers up to about
2.5~fm$^{-1}$~\cite{Baker:2023uzx,Chinn:1993zza},
and should be applied when comparing scattering calculations with the extracted local potential to
calculations based on the non-local Watson optical potential.

\begin{figure}
\begin{center}
\includegraphics[width=0.65\textwidth]{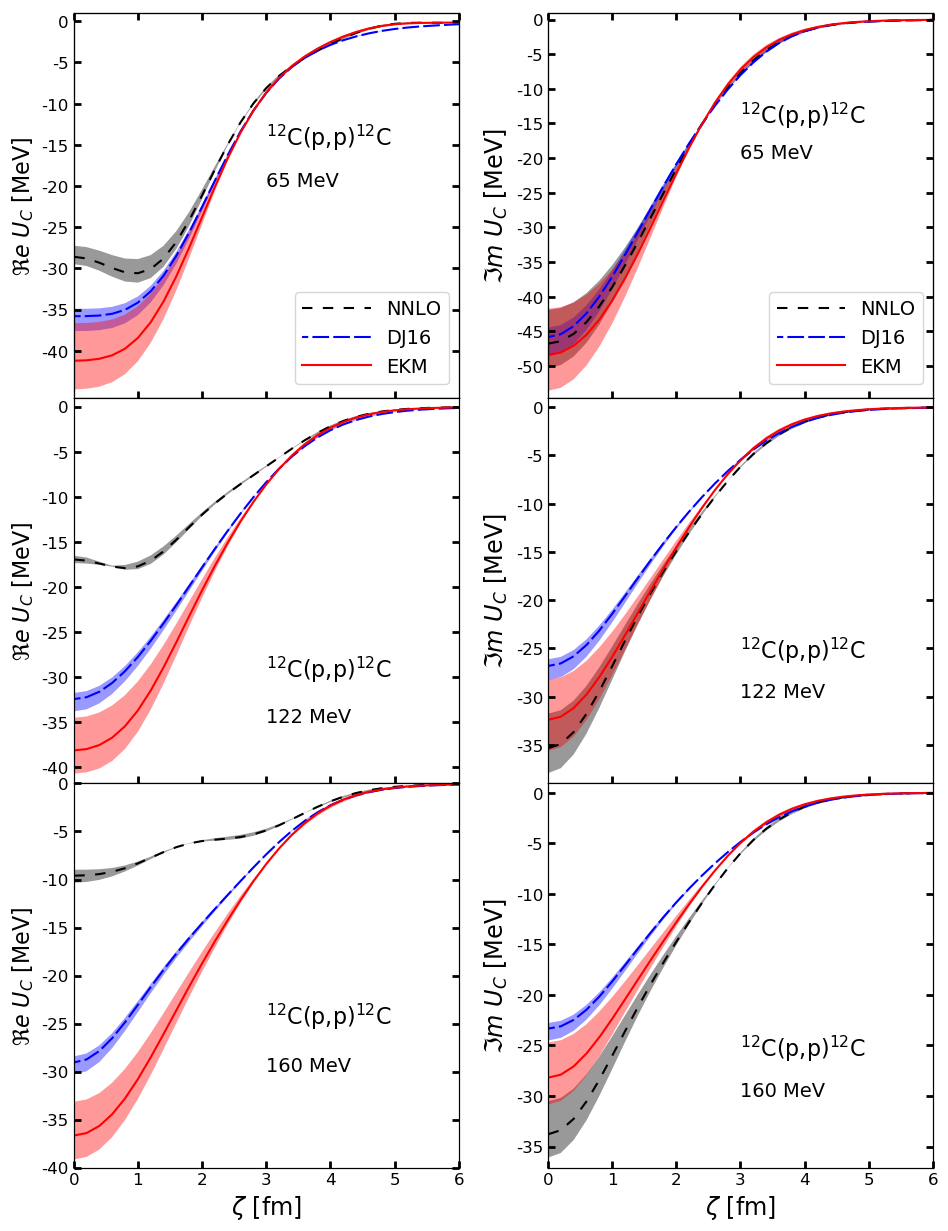}
\caption{The local piece of the effective central proton-$^{12}$C interaction $U_C$ as function of the
radial distance extracted at 65~MeV, 122 MeV,
and 160~MeV projectile energy based on the NNLO$_{\rm{opt}}$~\cite{Ekstrom13} chiral potential (short-dashed), the EKM chiral potential at order
N2LO~\cite{Epelbaum:2014sza,Epelbaum:2014efa} (solid), and the
Daejeon16 potential~\cite{Shirokov:2016ead} (dashed). The left panel depicts the real part of $U_C$, while
the right panel shows the imaginary part.  The description of the lines and bands is the same as in
Fig.~\ref{cs12C_log}. 
}
\label{fig11}
\end{center}
\end{figure}

The local effective central potentials  for proton scattering from $^{12}$C obtained this way
 are shown in Fig.~\ref{fig11}.    
First, we observe that all curves exhibit a Woods-Saxon shape, as it should be expected from 
the central part of local optical potentials. 
Next, we need to point out that one should not concentrate on differences in the potentials for $\zeta
\rightarrow 0$, since the coordinate space Schr\"odinger equation multiplies a factor $\zeta^2$ to the
potential. More important are differences for $\zeta$ between 1 and 3~fm.  
While at 65~MeV the three different potentials are quite similar in this range, at the higher energies
there is a striking difference between the potential extracted from NNLO$_{\rm opt}$ compared to the 
other two for the real central part.
The real central optical potential extracted from NNLO$_{\rm opt}$ becomes weaker as the energy
increases, which is consistent with the better description of the differential cross sections shown
in Fig.~\ref{cs12C_log}, and which can be traced back to a weaker central $NN$ part in NNLO$_{\rm opt}$ as shown
in Fig.~\ref{fig1}. The imaginary parts of the central optical potential do not exhibit a strong energy
dependence.

Obtaining the local part of the spin-orbit potentials in coordinate space is more involved. The momentum
space representation of the spin-orbit operator is given by $i(\bm{\sigma^{(0)}}\cdot\hat{\bm{n}})$,
where $\hat{\bm{n}}$ is a vector normal to the scattering plane, Eq.~(\ref{eq:2}), leading to a sine
function in the momentum space $\widehat{U}_{SO}$.  To obtain $U_{SO}(\zeta)$ we multiply with the KMT
factor and normalize the spin-orbit potential so that it is compatible with standard coordinate space
scattering codes~\cite{ThompsonNunes,Thompson:1988zz} that define the spin-orbit potential
as $\frac{2}{\zeta} U_{SO}(\zeta) (\bm{l} \cdot \bm{\sigma^{(0)}})$, where $\bm{l}$ is the orbital angular
momentum and $\bm{\sigma^{(0)}}$ the spin of the projectile. 

\begin{figure}
\begin{center}
\includegraphics[width=0.65\textwidth]{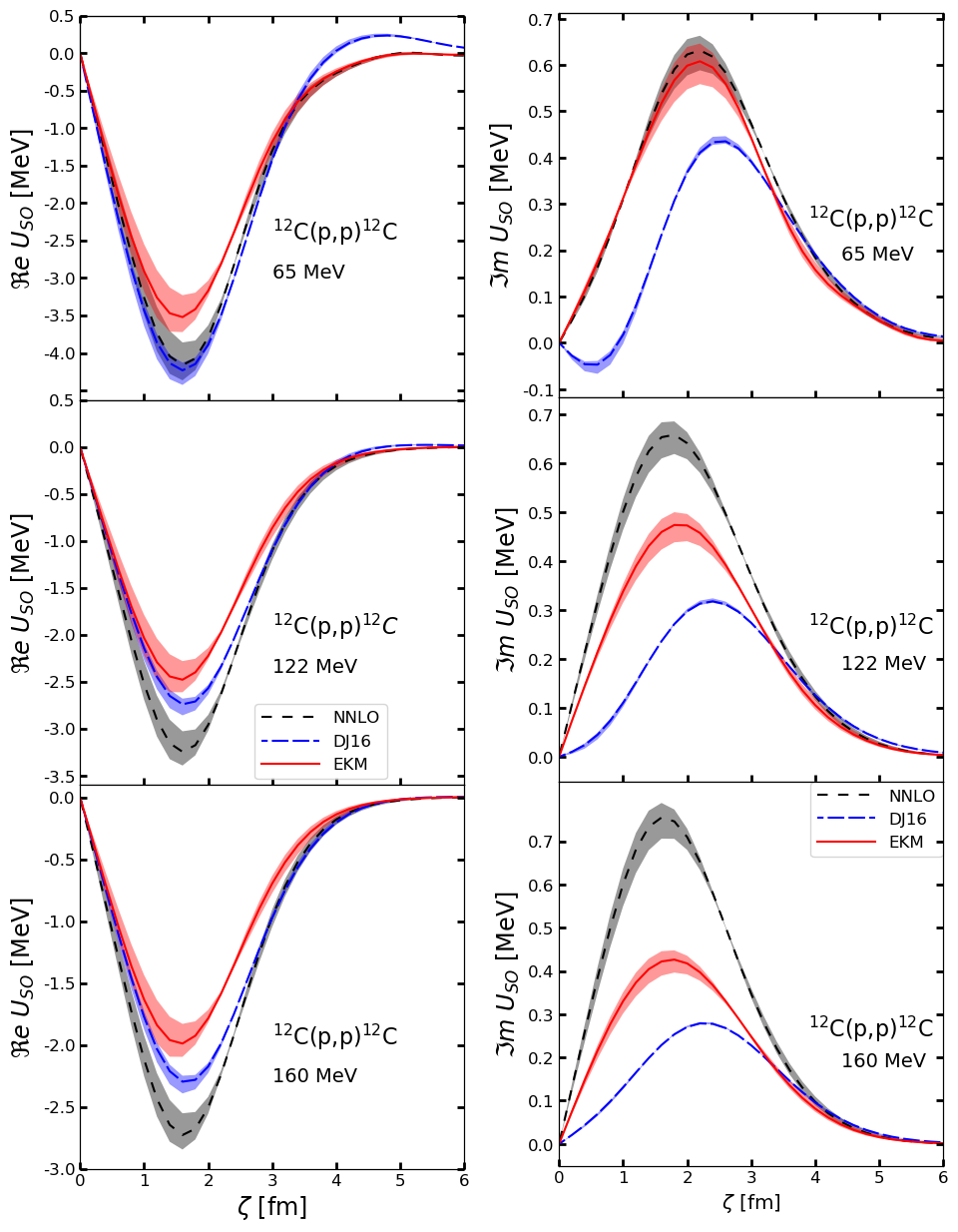}
\caption{The local piece of the effective spin-orbit proton-$^{12}$C interaction $U_{SO}$ as function of the
radial distance extracted at 65~MeV, 122 MeV,
and 160~MeV projectile energy based on the NNLO$_{\rm{opt}}$~\cite{Ekstrom13} chiral potential (short-dashed), the EKM chiral potential at order
N2LO~\cite{Epelbaum:2014sza,Epelbaum:2014efa} (solid), and the
Daejeon16 potential~\cite{Shirokov:2016ead} (dashed). The left panel depicts the real part of $U_{SO}$, while 
the right panel shows the imaginary part. The description of the lines and bands is the same as in
Fig.~\ref{cs12C_log}. Notice the difference in scale between the real and imaginary parts of the spin-orbit
interactions. 
}
\label{fig12}
\end{center}
\end{figure}

The local parts of the spin-orbit potentials based on the three different chiral $NN$ interactions are shown in
Fig.~\ref{fig12}. The real parts are relatively similar, with the NNLO$_{\rm opt}$ interactions being
the strongest. 
This is consistent with the $NN$ Wolfenstein amplitudes $C$ shown in Fig.~\ref{fig2}, where 
the NNLO$_{\rm opt}$ shows the largest spin-orbit contribution.
The imaginary spin-orbit potential is roughly an order of magnitude smaller than the real part.  In
some phenomenological optical potentials~\cite{Varner:1991zz} it was also found that the imaginary part
is quite small, and was thus omitted. Other phenomenological optical potentials~\cite{Weppner:2009qy} include
it, however the fit returns it as being quite small.

\begin{figure}
\begin{center}
\includegraphics[width=0.65\textwidth]{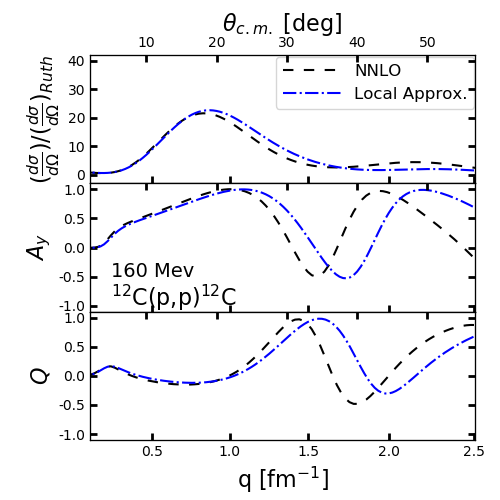}
\caption{The differential cross section divided by the Rutherford cross section (upper panel), the
analyzing power $A_y$ (middle panel), and the spin-rotation function $Q$ (bottom panel) for proton
scattering from $^{12}$C at 160~MeV. The dashed line represents the full {\it ab inito} calculation based
on the NNLO$_{\rm{opt}}$~\cite{Ekstrom13} chiral potential using $\hbar\omega=20$~MeV and $N_{\rm max}=10$,
while the dash-dotted line shows the calculation based only on the local part of the corresponding
effective interaction multiplied with the KMT factor. 
Details are given in the text. 
}
\label{fig14}
\end{center}
\end{figure}

The local parts of the effective potentials as shown in Fig.~\ref{fig11} exhibit for both the real and
imaginary pieces Woods-Saxon shapes. We picked the potentials based on $\hbar\omega$~=~20~MeV and were able
to fit them with 3 Woods-Saxon terms. We found that the spin-orbit term can also be very well represented by
Woods-Saxon terms multiplied by the radial variable $\zeta$. The details are given in the Appendix.   
To put this into perspective with calculations of $NA$ scattering observables, we then use the Woods-Saxon parameterization of the local part of the {\it ab initio}
effective interaction obtained from the NNLO$_{\rm opt}$ chiral interaction and calculate scattering
observables at 160~MeV using a coordinate space differential equation solver, and compare this result with the full
calculation (Fig.~\ref{fig14}). This comparison gives insight into the importance of the nonlocal structure
of the {\it ab initio} leading order $NA$ effective interaction. In all the observables, differential cross
section and spin observables, the differences are quite small at forward angles, i.e. small momentum
transfers. 
This indicates that at small momentum transfers the nonlocal structure contributes little, since here the
scattering can be viewed as a  grazing of the nucleus by the projectile.
However as the momentum transfer grows the projectile enters the target and non-local effects become important.


\section{Conclusions}
\label{sec:conclusions}

In this work we study structure and elastic scattering observables for light nuclei up the $p$-shell 
in an {\it ab initio} framework for three different $NN$ interactions based on chiral effective field
theory. We concentrate on the closed shell nuclei $^4$He and $^{16}$O, and the open-shell nucleus
$^{12}$C.  The elastic scattering observables are calculated in leading order in the spectator expansion.
This means that two nucleons are active in the scattering process, and no explicit three-nucleon forces
are included in the scattering process. Therefore, we concentrate on two chiral interactions up to
next-to-next-to-leading (NNLO) order, namely the NNLO$_{\rm{opt}}$ interaction from
Ref.~\cite{Ekstrom13}, and the EKM interaction from Refs.~\cite{Epelbaum:2014sza,Epelbaum:2014efa} up to
N2LO. The Daejeon16 interaction~\cite{Shirokov:2016ead} starts from the Idaho N3LO chiral
interaction~\cite{Entem:2003ft}, which is SRG evolved with a flow parameter
$\lambda$=1.5~fm$^{-1}$ and cast into a harmonic oscillator basis. The NNLO$_{\rm{opt}}$ chiral
interactions is fitted up to a $NN$ laboratory energy of about 125~MeV, while the other two interactions
give a good description of $NN$ to at least 200~MeV. We compare the $NN$ Wolfenstein amplitudes $A$ and
$C$ which are relevant for $NA$ scattering from nuclei with a 0$^+$ ground state to amplitudes
calculated from a `high-precision' $NN$ potential that can serve as representing $NN$ data, and see
that the NNLO$_{\rm{opt}}$ interaction provides a weaker central interaction at energies higher than
125~MeV. It also has a slightly stronger spin-orbit force. 

When considering structure observables, we focus on those that are dominated by one-body densities. The
latter enter the calculation of the leading order term of the spectator expansion, and thus may give a
connection between structure and scattering observables.  We presented ground state energies and
point-proton radii calculated with our choice of $NN$ interactions as well as the corresponding charge
form factors.  For the $p$-shell nuclei $^{12}$C and $^{16}$O the calculated charge form factors are
reasonably close to experiment, though the minima in both cases are located at slightly higher momentum
transfers, which should be related to the point-proton structure radius being predicted too small.
The calculated charge form factor for $^4$He is close to experiment up to about a momentum transfer
of 2~fm$^{-1}$, but completely misses the dip at about 3~fm$^{-1}$; the origin of this deficiency
is unclear but it could be an indication that the chiral expansion breaks down for larger momenta. 

The translationally invariant off-shell one-body densities together with the off-shell $NN$ Wolfenstein
amplitudes $A$, $C$, and $M$ enter the calculations for the leading order effective $NA$ interaction
for proton-nucleus elastic scattering observables are derived. For the $p$-shell nuclei $^{12}$C and
$^{16}$O we find that at the lowest projectile energy we consider, 65~MeV, all three chiral $NN$
interactions give relatively similar descriptions of the differential cross sections and under-predict
the first maximum, while at both higher energies, 100~MeV and 160 (180) MeV, they over-predict the
experiment. The NNLO$_{\rm{opt}}$ is closest to the experiment, though under-predicting  the $NN$ Wolfenstein
amplitudes. In order to gain insight into this apparent contradiction, we extracted the local part of
the effective $NA$ interaction and found that indeed the real part of the central potential calculated
from the NNLO$_{\rm{opt}}$ interaction is considerable weaker than those extracted from the other
interactions for energies higher than 100~MeV. The extracted imaginary parts of the central potential
are relatively similar as function of energy for the three chiral $NN$ interactions.  

Since the spin observables are calculated as products of real and imaginary parts of $NA$ Wolfenstein
amplitudes the large difference in the real central part between the NNLO$_{\rm{opt}}$ interaction 
and the EKM and Daejeon16 interaction propagates to the description of the analyzing power for $^{12}$C
and $^{16}$O. Here only the calculations based on the NNLO$_{\rm{opt}}$ chiral interaction reproduce
the shape of $A_y$ for energies larger than 100~MeV.   

The situation is slightly different for $^4$He, 
where for all energies considered the differential cross
section is overpredicted by the calculations when considering momentum transfers up to about
1.5-2.0~fm$^{-1}$, i.e. for the first maximum in the cross section. This momentum range 
is similar to the one for which the calculated charge form factor matches the corresponding data. Therefore,
one can not point to a single reason why the proton differential cross section is not well described
at higher momentum transfers. Though again the NNLO$_{\rm{opt}}$ chiral interaction gives the best
description from the three interactions. The analyzing powers shown at the same energies are not well
described either, although earlier work~\cite{Burrows:2020qvu} showed a very good description of 
differential cross section and analyzing power based on the NNLO$_{\rm{opt}}$ chiral interaction at 200~MeV
projectile energy. For the tightly bound $s$-wave dominated nucleus $^4$He it may very be the case that
higher orders in the spectator expansion are important for a better description of the lower energy
observables. 

Summarizing, our studies show that our choice of three $NN$ interactions that are based on a chiral
field theory framework and calibrated to reproduce similar data in the few-nucleon sector predict $NA$
scattering observables quite differently. Going into more detail by considering the calculated effective $NA$
interactions and extracting their local parts may be able to point to pieces in the $NN$
interactions that dominate their performance in $NA$ calculations.


\begin{acknowledgments}
This work was performed in part under the auspices of the U.S.~Department of Energy under contract Nos.~DE-FG02-93ER40756 and DE-SC0023495. The numerical computations benefited from computing resources provided by the National Energy Research Scientific Computing Center (NERSC), a U.~S.~DOE Office of Science User Facility located at Lawrence Berkeley National Laboratory, operated under contract No.~DE-AC02-05CH11231.
\end{acknowledgments}

\clearpage

\newpage

\begin{appendix}

\section{Woods-Saxon parameterization of the local parts of the effective $NA$ interactions for $^{12}$C}

The parameterization of local optical potentials in terms of Woods-Saxon shapes has a long history.
Specifically, code packages like FRESCO~\cite{Thompson:1988zz,FRESCO} and ECIS~\cite{ECIS,ECIS-12} 
allow to enter parameters based on Woods-Saxon parameterization to be entered for calculating elastic $NA$
scattering observables. In order to investigate if the local parts of the effective interactions for $^{12}$C
we calculated {\it ab initio} follow standard expectations of having Woods-Saxon shapes, 
we list
the coefficients for a local coordinate space parts of the optical potentials for $^{12}$C 
featured in this work at three different proton projectile energies. We choose the calculations
based on $\hbar\omega=20$ MeV and $N_{\rm max}=10$  from Figs.~\ref{fig11} and~\ref{fig12}.
We find that three Woods-Saxon shapes are sufficient in all cases to represent the local parts of the
effective interaction, when setting as criteria 
that the analytical function correlate
with the numerical results at high precision ($>$ 0.9999). 
Figure~\ref{fig14} of the main text  shows as dashed line a scattering calculation which featuring the 
NNLO$_{\rm{opt}}$ chiral potential from Ref.~\cite{Ekstrom13} with a proton projectile at 160 MeV.  
The dashed-dotted
line is calculated using the optical potential based on the coefficients listed in this appendix 
which are processed through an ECIS type  r-space scattering code. 

\subsection{Parameters for the central part of the extracted local optical potential}

The central part of the optical potentials is described by three Woods-Saxon terms for the real and imaginary part:
\begin{eqnarray}
	\Re e \: U_c (\zeta)& =& \sum_{i=1}^3 V_i \frac{1}{1+e^{\left( \frac{\zeta-R_i}{a_i}\right)}} \cr
	\Im m \: U_c (\zeta)& =& \sum_{i=1}^3 W_i \frac{1}{1+e^{\left( \frac{\zeta-S_i}{b_i}\right)}}
\label{eq:central}
\end{eqnarray}
Three different energies are fitted separately, all the central potentials form the traditional negative energy nuclear well.

\begin{table}[!hb]
\begin{center}
\caption{Woods-Saxon parameters for $\Re e \: U_c (\zeta)$ for $^{12}$C based on the
NNLO${_{\rm{opt}}}$~\cite{Ekstrom13} chiral $NN$ potential at the energies indicated.}
\begin{tabular}{|l|c|c|c|}
\hline
    & 65 MeV & 122 MeV & 160 MeV \\
\hline \hline
V$_1$  & -2.9585 & -9.0907 & -7.4562   \\ 
V$_2$  & -35.8338 & -11.6254  & -3.8378   \\
V$_3$  & 10.3808 &  3.8241 & 1.5649   \\
R$_1$  & 4.1455 &  3.3838  & 3.4214  \\
R$_2$  &2.0327 &   1.6355   & 1.2032   \\
R$_3$  & 0.7625 & 0.6023   & 2.4738   \\
a$_1$  & 0.3194 & 0.5202  & 0.5283   \\
a$_2$  & 0.5487 & 0.4387  & 0.3217  \\
a$_3$  & 0.3402  & 0.2591  & 0.3207  \\
\hline\hline
\end{tabular}
\end{center}
\end{table}
\begin{table}[!hb]
\begin{center}
\caption{Woods-Saxon parameters for $\Re e \: U_c (\zeta)$ for $^{12}$C based on the
EKM~\cite{Epelbaum:2014sza,Epelbaum:2014efa} chiral $NN$ potential up to N2LO at the energies indicated.}
\begin{tabular}{|l|c|c|c|}
\hline
    & 65 MeV & 122 MeV & 160 MeV \\
\hline \hline
V$_1$  & -1.2962 & -16.6887 & -16.8768   \\
V$_2$  & -49.5527 & -15.5814  & -14.8252   \\
V$_3$  & 9.4976 &  -6.3532 & -5.6641   \\
R$_1$  & 4.1469 &  2.9288  & 2.9069   \\
R$_2$  & 1.9121 &   1.8347   & 1.6688   \\
R$_3$  & 0.6190 & 1.1140   & 0.8959   \\
a$_1$  & 0.2593 & 0.5773  & 0.5825   \\
a$_2$  & 0.6276 & 0.4057  & 0.4322  \\
a$_3$  & 0.4956  & 0.2870  & 0.2696  \\
\hline\hline
\end{tabular}
\end{center}
\end{table}
\begin{table}[!hb]
\begin{center}
\caption{Woods-Saxon parameters for $\Re e \: U_c (\zeta)$ for $^{12}$C based on the Daejeon16~\cite{Shirokov:2016ead} $NN$ potential at the energies indicated.}
\begin{tabular}{|l|c|c|c|}
\hline
    & 65 MeV & 122 MeV & 160 MeV \\
\hline \hline
V$_1$  & -3.2707 & -19.1730 & -14.9108   \\
V$_2$  & -37.4844 & -10.4265  & -9.8969   \\
V$_3$  & 4.6176 &  -3.5101 & -4.9593   \\
R$_1$  & 3.8250 &  2.7855  & 2.9483   \\
R$_2$  & 2.0646 &   1.6311   & 1.4607   \\
R$_3$  & 0.5782 & 0.8474   & 0.7528   \\
a$_1$  & 0.9023 & 0.6543  & 0.6082   \\
a$_2$  & 0.5911 & 0.4200  & 0.4283  \\
a$_3$  & 0.3803  & 0.2468  & 0.2413  \\
\hline\hline
\end{tabular}
\end{center}
\end{table}
\begin{table}[!hb]
\begin{center}
\caption{Woods-Saxon parameters for $\Im m \: U_c (\zeta)$ for $^{12}$C based on the NNLO${_{\rm{opt}}}$ $NN$ potential at the energies indicated.}
\begin{tabular}{|l|c|c|c|}
\hline
    & 65 MeV & 122 MeV & 160 MeV \\
\hline \hline
W$_1$  & -19.3663  & -14.2327  & -13.7046   \\
W$_2$  & -22.2666 & -15.3673     & -14.3915 \\
W$_3$  & -6.0634 & -6.6642   & -6.5563   \\
S$_1$  &  1.5788 & 1.4190   & 1.4850   \\
S$_2$  & 2.5801 & 2.7300    & 2.7570   \\
S$_3$  & 0.8166 & 0.7380     & 0.7570   \\
b$_1$  & 0.4035 & 0.4105    & 0.4262   \\
b$_2$  & 0.5590 & 0.5645   & 0.5462  \\
b$_3$  & 0.2433 & 0.2346   & 0.2436 \\
\hline\hline
\end{tabular}
\end{center}
\end{table}
\begin{table}[!hb]
\begin{center}
\caption{Woods-Saxon parameters for $\Im m \: U_c (\zeta)$ for $^{12}$C based on the EKM $NN$ potential at the energies indicated.}
\begin{tabular}{|l|c|c|c|}
\hline
    & 65 MeV & 122 MeV & 160 MeV \\
\hline \hline
W$_1$  & -15.7595  & -12.6308  & -11.0227   \\
W$_2$  & -28.7883 & -15.6105     & -13.1028 \\
W$_3$  & -4.7952 & -4.8887   & -4.7258   \\
S$_1$  &  1.5426 & 1.5055   & 1.5297   \\
S$_2$  & 2.3406 & 2.6002    & 2.6623   \\
S$_3$  & 0.8233 & 0.7712     & 0.7698   \\
b$_1$  & 0.3747 & 0.4097    & 0.4253   \\
b$_2$  & 0.5709 & 0.5727   & 0.5625  \\
b$_3$  & 0.2270 & 0.2376   & 0.2436 \\
\hline\hline
\end{tabular}
\end{center}
\end{table}
\begin{table}[!hb]
\begin{center}
\caption{Woods-Saxon parameters for $\Im m \: U_c (\zeta)$ for $^{12}$C based on the Daejeon16 $NN$ potential at the energies indicated.}
\begin{tabular}{|l|c|c|c|}
\hline
    & 65 MeV & 122 MeV & 160 MeV \\
\hline \hline
W$_1$  & -21.5698  & -11.6742  & -11.2472   \\
W$_2$  & -17.7607 & -11.0697     & -7.9812 \\
W$_3$  & -7.4589 & -4.7278   & -4.7165   \\
S$_1$  &  1.6083 & 1.5843   & 1.6829   \\
S$_2$  & 2.8052 & 2.8878    & 3.0711   \\
S$_3$  & 0.8177 & 0.8136     & 0.8578   \\
b$_1$  & 0.4242 & 0.4419    & 0.4714   \\
b$_2$  & 0.5218 & 0.5893   & 0.5571  \\
b$_3$  & 0.2544 & 0.2561   & 0.2748 \\
\hline\hline
\end{tabular}
\end{center}
\end{table}


\subsection{Parameters for the spin-orbit part of the extracted local optical potential}

For the Spin-Orbit potential, the following functional forms were most appropriate for our data:
\begin{eqnarray}
	\Re e \: U_{SO} (\zeta)& =& \sum_{i} Vso_i \frac{\zeta}{1+e^{\left( \frac{\zeta-Rso_i}{aso_i}\right)}} \cr
	\Im m \: U_{SO} (\zeta)& =& \sum_{i} Wso_i \frac{\zeta}{1+e^{\left( \frac{\zeta-Sso_i}{bso_i}\right)}}
\label{eq:spin-orbit}
\end{eqnarray}
Here we varied the amount of Woods-Saxon terms needed, our objective again was to  obtaining an excellent 
correlation ($>$ 0.9999).
Since many r-space codes (like FRESCO~\cite{Thompson:1988zz,FRESCO,ThompsonNunes} 
for example) add an obligatory $\frac{2}{\zeta}$ to their spin orbit term. The complete
term run through the r-space scattering code is
\begin{equation}
	{\cal U}_{SO} = \frac{2}{\zeta}\:U_{SO} (\zeta)\:(2{\vec l}\cdot{\vec \sigma}),
\end{equation}	
where our contribution, $U_{SO} (\zeta)$, is defined in Eq.~\ref{eq:spin-orbit}. The second factor of $2$, 
included with the vector operator, is tradition within most r-space scattering codes.
We found in all the cases listed here that the real spin-orbit formed a large attractive nuclear well while the imaginary
term formed a smaller repulsive well. These results are in line with most phenomenological results.

\begin{table}
\begin{center}
	\caption{Woods-Saxon parameters for $\Re e \: U_{SO} (\zeta)$ for $^{12}$C based on the NNLO${_{\rm{opt}}}$ $NN$ potential at the energies indicated.}
\begin{tabular}{|l|c|c|c|}
\hline
    & 65 MeV & 122 MeV & 160 MeV \\
\hline \hline
Vso$_1$  & -3.2704 & -1.8966  &  -2.3383 \\
Vso$_2$  & -0.2579 &  -0.7528 &  0.0 \\
Rso$_1$  & 2.1307 & 2.3350 & 2.0813 \\
Rso$_2$  & 1.5961 & 1.7201  & \\
aso$_1$  & 0.4663 &0.4610 & 0.4946 \\
aso$_2$  & 0.2043 & 0.2681 & \\
\hline\hline
\end{tabular}
\end{center}
\end{table}
\begin{table}
\begin{center}
        \caption{Woods-Saxon parameters for $\Re e \: U_{SO} (\zeta)$ for $^{12}$C based on the EKM $NN$ potential at the energies indicated.}
\begin{tabular}{|l|c|c|c|}
\hline
    & 65 MeV & 122 MeV & 160 MeV \\
\hline \hline
Vso$_1$  & -1.7713 & -2.3492  &  -1.8878 \\
Vso$_2$  & -1.6051  &  0.0 &  0.0 \\
Rso$_1$  & 2.2809 & 1.9516 & 1.9512 \\
Rso$_2$  &  1.5784&   & \\
aso$_1$  & 0.4889 &0.5267 & 0.5275 \\
aso$_2$  & 0.4493 & & \\
\hline\hline
\end{tabular}
\end{center}
\end{table}
\begin{table}
\begin{center}
        \caption{Woods-Saxon parameters for $\Re e \: U_{SO} (\zeta)$ for $^{12}$C based on the Daejeon16 $NN$ potential at the energies indicated.}
\begin{tabular}{|l|c|c|c|}
\hline
    & 65 MeV & 122 MeV & 160 MeV \\
\hline \hline
Vso$_1$  & -2.4902 & -1.6768  &  -1.4797 \\
Vso$_2$  & -1.5978  & -0.6924 &  -0.4470 \\
Vso$_3$  & 0.1368   & 0.03618   & 0.0     \\
Rso$_1$  & 2.3504 & 2.3691 & 1.9384 \\
Rso$_2$  & 1.5573 & 1.6801  & 2.9354\\
Rso$_3$  & 4.8284 & 4.7296  &  \\
aso$_1$  & 0.5137 &0.5229 & 0.4377 \\
aso$_2$  & 0.4405 & 0.3417 & 0.4365\\
aso$_3$  & 0.5732 & 0.6223& \\
\hline\hline
\end{tabular}
\end{center}
\end{table}
\begin{table}
\begin{center}
	\caption{Woods-Saxon parameters for $\Im m \: U_{SO} (\zeta)$ for $^{12}$C based on the NNLO${_{\rm{opt}}}$ $NN$ potential at the energies indicated.}
\begin{tabular}{|l|c|c|c|}
\hline
    & 65 MeV & 122 MeV & 160 MeV \\
\hline \hline
Wso$_1$  & 0.4369  & 0.4194 & 0.7072  \\
Wso$_2$  & -0.2054 & 0.1387 & 0.0 \\
Sso$_1$  & 2.6510 & 1.9424 & 2.013 \\
Sso$_2$  & 0.9638 & 3.1285  &  \\
bso$_1$  & 0.6273 & 0.5067 & 0.5988 \\
bso$_2$  & 0.3738 & 0.5462 &  \\
\hline\hline
\end{tabular}
\end{center}
\end{table}
\begin{table}
\begin{center}
        \caption{Woods-Saxon parameters for $\Im m \: U_{SO} (\zeta)$ for $^{12}$C based on the EKM $NN$ potential at the energies indicated.}
\begin{tabular}{|l|c|c|c|}
\hline
    & 65 MeV & 122 MeV & 160 MeV \\
\hline \hline
Wso$_1$  & 0.8261  & 0.3767 & 0.3843  \\
Wso$_2$  & -0.5784 & 0.0 & 0.0 \\
Sso$_1$  & 2.3345 & 2.3452 & 2.1325 \\
Sso$_2$  & 1.6682 &   &  \\
bso$_1$  & 0.6391 & 0.6396 & 0.6540 \\
bso$_2$  & 0.6645 &  &  \\
\hline\hline
\end{tabular}
\end{center}
\end{table}
\begin{table}
\begin{center}
        \caption{Woods-Saxon parameters for $\Im m \: U_{SO} (\zeta)$ for $^{12}$C based on the Daejeon16 $NN$ potential at the energies indicated.}
\begin{tabular}{|l|c|c|c|}
\hline
    & 65 MeV & 122 MeV & 160 MeV \\
\hline \hline
Wso$_1$  & 0.3018  & 0.1932 & 0.1664  \\
Wso$_2$  & -0.4981 & -0.1558 & -0.0609 \\
Sso$_1$  & 2.8836 & 2.9440 & 2.9051 \\
Sso$_2$  & 1.0721 & 0.9685  &0.9725  \\
bso$_1$  & 0.6643 & 0.6404& 0.6048 \\
bso$_2$  & 0.4454 & 0.3715 & 0.2847 \\
\hline\hline
\end{tabular}
\end{center}
\end{table}

\end{appendix}
\clearpage

\bibliography{denspot,clusterpot,ncsm,reactions}

\clearpage

\clearpage

\end{document}